\documentclass[fleqn,usenatbib]{mnras}
\usepackage{mathtools}
\usepackage{enumitem}
\usepackage{appendix}
\usepackage{xcolor}
\usepackage{subfigure}
\usepackage{graphicx}
\usepackage{amsmath}
\usepackage{figsize}
\usepackage{txfonts}
\usepackage[T1]{fontenc}
\usepackage{threeparttable}
\usepackage{CJKutf8}

\DeclareRobustCommand{\VAN}[3]{#2}
\let\VANthebibliography\thebibliography
\def\thebibliography{\DeclareRobustCommand{\VAN}[3]{##3}\VANthebibliography}

\newcommand{\black}[1]{\textcolor{black}{#1}}

\newcommand{\tnr}[2]{{#1}_\mathrm{#2}}

\title[Synchrotron Radiation from the Dragonfly Galaxy]{Revisiting the Dragonfly Galaxy I. High-resolution ALMA and VLA Observations of the Radio Hotspots in a Hyper-luminous Infrared Galaxy at $z=1.92$}

\author[Yuxing Zhong et al.]{Yuxing Zhong$^{1}$ \begin{CJK*}{UTF8}{gkai}(仲宇星)\end{CJK*},
Akio K. Inoue$^{1,2}$,
Yuma Sugahara$^{2,3}$,
Kana Morokuma-Matsui$^{4}$,
Shinya Komugi$^{5}$,
\newauthor{Hiroyuki Kaneko$^{6,7,3}$,
and Yoshinobu Fudamoto$^{2,3}$}
\\
$^{1}$Department of Pure and Applied Physics, Waseda University, 3-4-1 Okubo, Shinjuku, Tokyo 169-8555, Japan\\
$^{2}$Waseda Research Institute of Science and Engineering, Waseda University, 3-4-1, Okubo, Shinjuku, Tokyo 169-8555, Japan\\
\black{$^{3}$National Astronomical Observatory of Japan, 2-21-1 Osawa, Mitaka, Tokyo 181--8588, Japan}\\
\black{$^{4}$Center for Computational Sciences, University of Tsukuba, Ten-nodai, 1-1-1 Tsukuba, Ibaraki 305-8577, Japan}\\
\black{$^{5}$Department of Liberal Arts, Kogakuin University, 2665-1 Nakano-cho, Hachioji, Tokyo 192-0015, Japan}\\
\black{$^{6}$Joetsu University of Education, 1, Yamayashiki-machi, Joetsu, Niigata 943--8512, Japan}\\
\black{$^{7}$Ibaraki University, 2-1-1 Bunkyo, Mito, Ibaraki 310-8512, Japan}
}

\date{Accepted XXX. Received YYY; in original form ZZZ}

\pubyear{2023}

\begin{document}
\label{firstpage}
\pagerange{\pageref{firstpage}--\pageref{lastpage}}
\maketitle

\begin{abstract}
Radio-loud active galactic nuclei (RLAGNs) are rare among AGN populations.
Lacking high-resolution and high-frequency observations, their structure and evolution stages are not well understood at high redshifts.
In this work, we report ALMA 237 GHz continuum observation at $0\farcs023$ resolution and VLA 44 GHz continuum observation at $0\farcs08$ resolution of the radio continuum emission from a high-redshift radio and hyper-luminous infrared galaxy at $z=1.92$.
The new observations confirm the South-East (SE) and North-West (NW) hotspots identified by previous low-resolution VLA observations at 4.7 and 8.2 GHz and identify a radio core undetected in all previous observations.
The SE hotspot has a higher flux density than the NW one does by a factor of 6, suggesting that there can be a Doppler boosting effect in the SE one.
In this scenario, we estimate the advance speed of the jet head, ranging from $\sim$0.1c -- 0.3c, which yields a mildly relativistic case.
The projected linear distance between the two hotspots is $\sim13$ kpc, yielding a linear size ($\leq20$ kpc) of a Compact-Steep-Spectrum (CSS) source. 
Combined with new \black{high-frequency ($\tnr{\nu}{obs}\geq44$ GHz) and archived low-frequency observations ($\tnr{\nu}{obs}\leq8.2$ GHz)}, we find that injection spectra of both NW and SE hotspots can be fitted with a continuous injection (CI) model.
Based on the CI model, the synchrotron ages of NW and SE hotspots have an order of $10^5$ yr, consistent with the order of magnitude $10^3 - 10^5$ yr observed in CSS sources associated with radio AGNs at an early evolution stage.
The CI model also favors the scenario in which the double hotspots have experienced a quiescent phase, suggesting that this RLAGN may have transient or intermittent activities.
\end{abstract}


\section{Introduction}\label{intro}
Radio galaxies hosting radio-loud active galactic nuclei (RLAGNs, $L\mathrm{_{6GHz}\gtrsim10^{23.2}\ erg\ s^{-1}}$; \citealt{2016ApJ...831..168K}) have their radio emissions dominated by magnetobremsstrahlung, i.e., synchrotron radiation attributed to the central AGN.
RLAGNs are rare amongst all AGN populations ($15-20\%$; \citealt{1989AJ.....98.1195K,2018MNRAS.475.3429W}).
In the local Universe, RLAGNs are found to preferentially reside in early-type galaxies with high stellar masses, low star formation rates, and radiatively inefficient AGNs \citep{2005MNRAS.362...25B,2014ARA&A..52..589H,2015MNRAS.452.3776G}.
At higher redshifts, there is an increasing number of RLAGNs related to active supermassive black holes (SMBHs) showing high accretion rates, and these AGNs are intrinsically radiatively efficient \citep{2018MNRAS.475.3429W,2020NewAR..8801539H}.
This different feature provides a good channel to access a more complete picture of the AGN evolution itself, as well as the co-evolution with its host galaxy.

To select radio AGNs hosted by high redshift radio galaxies (HzRGs), a feature called Ultra-Steep-Spectrum (USS), with a spectral index $\alpha\leq-1.3$ ($S_\nu\propto\nu^\alpha$) in the low-frequency radio spectral energy distribution (SED) has been proven an effective technique \citep{2000A&AS..143..303D}.
Such a steeper slope at high redshifts is considered a result of two possible causes.
The first one is the Malmquist effect, due to which the cut-off frequency may migrate towards lower frequencies since it is anti-correlated with the radio luminosity.
The second possibility is that HzRGs may reside in denser environments because of the evolution of cosmic neutral hydrogen density, leading to a steeper initial power-law of the energy distribution \citep{2008A&ARv..15...67M}.
Recent studies have emphasized the importance of cosmic microwave background (CMB) radiation \citep[e.g.,][]{2017MNRAS.469.4083S}.
The inverse Compton scattering that involves CMB photons and synchrotron-radiating electrons could be the dominant energy loss and result in a rapid decrease in the energy of synchrotron-emitting relativistic electrons.

Other than USS, many HzRGs show Peaked-Spectrum (PS) or Compact-Steep-Spectrum (CSS) \citep{1997AJ....113..148O}.
PS sources always have a frequency at which the flux density peaks.
This frequency is known as peaked or turnover frequency, below which the medium is optically-thick for the synchrotron-emitting source due to the synchrotron self-absorption (SSA) and/or free-free absorption (FFA), showing an inverse spectral index \citep{2016era..book.....C,snell2019fundamentals}.
These sources are considered the youngest populations of radio AGNs.
The PS sources may evolve into CSS sources with evolving jets, characterized by a typical projected linear size $\leq20$ kpc and $\alpha\leq-0.5$ at low frequencies \citep{2021A&ARv..29....3O}.

Many high redshift CSS sources have been observed at low frequencies over the decades, while the high-frequency observations are still lacking \citep{2000AAS..145..121P,2020ApJ...896...18P,2019AstBu..74..348S}.
The high-frequency behavior of the synchrotron radiation is crucial to investigate the evolution of the radio AGN, especially when the radio jets launched by the AGN have interacted with their ambient medium.
Radio hotspots that are working surfaces where the radio jet has interacted with the medium are essential to answer this question.
The synchrotron age of the radio hotspot is determined by the break frequency above which the spectrum steepens, together with the magnetic field in equipartition.
This break frequency can be estimated by fitting the observed radio spectrum covering a range of frequencies from low to high to an injection spectrum of the accelerated synchrotron-emitting electrons.

In this work, we have studied a HzRG -- MRC 0152-209, named Dragonfly galaxy, which has a star formation rate (SFR) of $\mathrm{\sim3000\ M_\odot\ yr^{-1}}$ and $M_\star\sim\mathrm{5\times10^{11}\ M_\odot}$ \citep{2015A&A...584A..99E,2015MNRAS.451.1025E}.
This galaxy constitutes of three components: the North-West (NW) galaxy, the South-East (SE) galaxy, and a Companion (C) with massive molecular gas (see panels (a), (NW), and (SE) in Fig.~\ref{fig:imaging}). 
The low-frequency observations at 4.7 and 8.2 GHz using the Very Large Array (VLA) were studied by \citet{2000AAS..145..121P}, and they identified two radio hotspots, the NW and SE ones, while the radio core remained non-detection (see green contours in panel (a) in Fig.~\ref{fig:imaging}).
New VLA Band Q observations (Project ID: VLA/15A-316 and VLA/17B-444, PI: Bjorn Emonts), combined with high-angular resolution Atacama Large Millimeter/submillimeter Array (ALMA) Cycle 4 (Project ID: 2016.1.01417.S, PI: Bjorn Emonts) and 6 (Project ID: 2018.1.00293.S, PI: Bjorn Emonts) observations in Band 6, allow us to have a deeper investigation into the radio properties including the lifetime of hotspots and the geometry of radio jets.

Throughout this paper, we assume a $\mathrm{\Lambda CDM}$ cosmology with $\Omega_m=0.309$, $\Omega_\Lambda=0.691$, and $H_0=\mathrm{67.7\ km\ s^{-1}\ Mpc^{-1}}$ \citep{2016A&A...594A..13P}.
Based on these assumptions, the luminosity distance of the Dragonfly galaxy is $\sim$15200 Mpc, and $1\arcsec$ corresponds to a projected physical scale of 8.62 kpc.\footnote{The calculation has made use of \citet{2006PASP..118.1711W}.}

\begin{table*}
\caption{Summary of the Observations}
\label{tab:obs}
\begin{tabular}{ccccccc}
\hline
Observation & Date & Frequency & Band & beam size & $\sigma\ (\mathrm{mJy\ beam^{-1}})$ \\
\hline
VLA A-configuration & 5th and 13th August, 2015 & 44 GHz & Band Q & $0\farcs07\times0\farcs04$ & 0.01 \\
VLA BnA-configuration & 29th May, 2015 & 44 GHz & Band Q & $0\farcs14\times0\farcs08$ & 0.018 \\
VLA B-configuration & 30th December, 2017 & 44 GHz & Band Q & $0\farcs26\times0\farcs13$ & 0.024 \\
ALMA Cycle 4 & 9 and 17th August, 2017 & 237 GHz & Band 6 & $0\farcs11\times0\farcs08$ & 0.011 \\
ALMA Cycle 6 & 23th June, 2019 & 237 GHz & Band 6 & $0\farcs026\times0\farcs023$ & 0.0065 \\
\hline
\end{tabular}
\end{table*}

\section{Observations}
In this section, we briefly describe the observational data used in this work, including observation setups, calibration procedures, and imaging strategies.
The observation dates, frequencies, bands, restored beam sizes, and root-mean-squares (RMS, $\sigma$) are summarized in Table.~\ref{tab:obs}.

\subsection{VLA Band Q}
The VLA observations were conducted in B-, BnA-, and A-configurations.
All the observations were centred at 44 GHz with an effective bandwidth of 7.5 GHz.
The total on-source time is 42 min for B- and BnA-configuration observations, and 178 min for A-configuration.
For all the observations, calibrator J2253+1608 was chosen to calibrate the bandpass response, calibrator J0204-1701, which is separated from the Dragonfly galaxy by $4.4^\circ$, was chosen to calibrate complex gain, including amplitude and phase, between the scans of the science target, and 3C147 was chosen to calibrate the flux density scale.

The B-configuration observation was calibrated by means of running the VLA calibration script that is provided by the National Radio Astronomy Observatory (NRAO) on the Common Astronomy Software Applications (CASA) package version 6.2.1 embedding the VLA calibration pipeline \citep{2007ASPC..376..127M,2022arXiv221002276T}.
The BnA- and A-configuration observations were calibrated by requesting pipeline calibrations through the NRAO Science Helpdesk.
All the imaging procedures were performed on CASA version 6.4.0.
For all the observations, using the 'hogbom' deconvolution algorithm, we imaged the data choosing the 'briggs' weighting with a robustness parameter +0.5, and put a mask on the strongest signal to avoid artefacts.
Due to the low signal-to-noise ratio of each spectral window, no self-calibration in any dataset can be applied, leaving low-level sidelobe contamination on the clean images.

\subsection{ALMA Band 6}
The ALMA Cycle 4 observations were performed on 9 and 17 August 2017 for 1.2 hours of on-source time with 45 antennas and baselines up to $\sim3.6$ km.
The ALMA Cycle 6 observations were conducted on 23 June 2019 for 2.2 hours of on-source time with 48 antennas and the longest baseline is up to 11.5 km.
For both observations, there are four spectral windows configured to cover two 4 GHz bands, one of which includes $235.83-239.58$ GHz to observe the redshifted CO(6-5) line emission $(\nu\mathrm{_{rest}\approx691.47\ GHz})$ and another includes $251.20-255.95$ GHz such that only continuum is observed.
The data calibrations were performed via the ALMA pipeline that is included in CASA version 4.7.2 for Cycle 4 and 5.4.0 for Cycle 6, respectively, by running the calibration script supplied with the data by the North American ALMA Science Center (NAASC).

The main objective of these observations is to image the CO(6-5) line emission, which is beyond the scope of this paper.
An analysis of CO(6-5) line emission based on Cycle 4 data is discussed in Lebowitz et al. (2023) and another analysis of CO(6-5) using combined data from Cycles 4 and 6 will be presented in a forthcoming paper (Paper II; Zhong et al. in prep.)
The detailed analysis of CO(6-5) line emission will be presented in a forthcoming paper (Paper II; Zhong et al. in prep.).
We adopt a uniform method for Cycle 4 and 6 observations to image the continuum in this work.
Prior to imaging the continuum, we flagged the channels with CO(6-5) line emission, as well as pseudo-lines due to the severe atmospheric absorption.
Next, we created a dirty image without any clean to calculate the root-mean-square noise $(\sigma)$ under the 'briggs' weighting with a robustness parameter +0.5.
Then, we cleaned the image non-interactively by setting $3\sigma$ as the stop threshold of the cleaning.

\subsection{Alignment}
\black{
The imaging of ALMA Cycle 4 and 6 and VLA 44 GHz observation in three configurations is the direct 'tclean' product of the measurement sets that are calibrated from raw files using the calibration
pipeline provided by NRAO or under the assistance of the NRAO and EA ALMA staff. 
No postprocessing, including self-calibration and manipulations on the WCS of the imaging results, was performed. 
Then the contours from different observations are overlaid on false-color images based on their WCS using Cube Analysis and Rendering Tool for Astronomy \citep[CARTA,][]{angus_comrie_2021_4905459}.
VLA 8.2 GHz observation is a reproduction of \citet{2000AAS..145..121P}. 
It was self-calibrated, and thus its WCS cannot be trusted\footnote{See Post-processing Section at \url{https://science.nrao.edu/facilities/vla/docs/manuals/oss2015B/performance/positional-accuracy}}. 
We, therefore, take the peak pixel of the SE hot spot observed in VLA A-configuration at 44 GHz as the reference to shift the WCS of VLA 8.2
GHz such that they match.}

\black{ Taking the SE peak pixel as the reference, the ALMA Cycle 6 has an offset of $0\farcs025$ between VLA 44 GHz in BnA-configuration along RA, and an offset of $0\farcs011$ between VLA 44 GHz in A-configuration along RA. Offsets along the DEC are completely negligible. 
Although these offsets
can be larger than the typical position uncertainty of one-tenth of beam size since there can be low-level systematic uncertainties, the current alignments are sufficient to suggest that the features originate
from the same radio source.}

\begin{figure*}
\begin{center}
\includegraphics[width=1.0\textwidth]{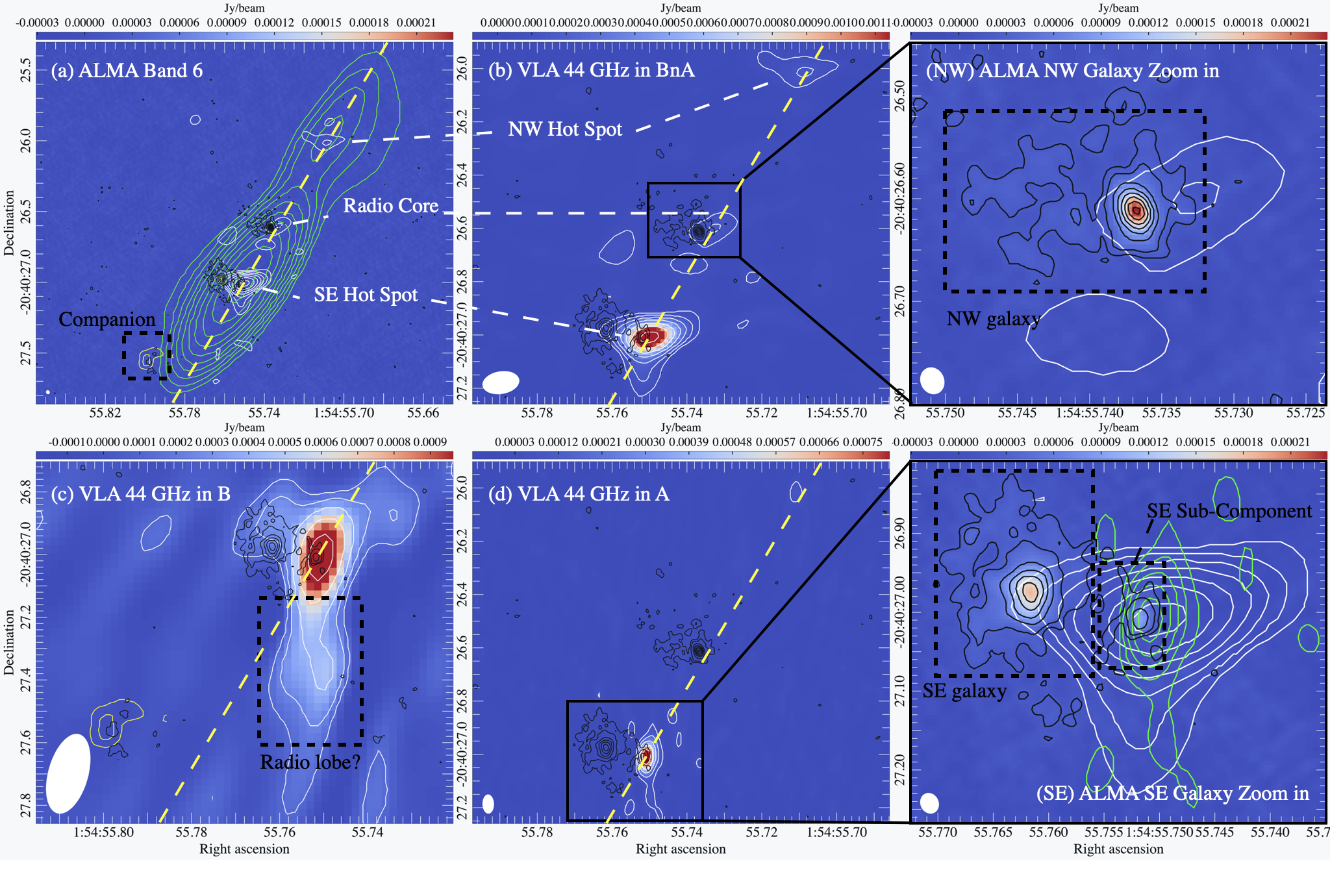}
\caption{(a) ALMA Cycle 6 Band 6 continuum image $\mathrm{(\sigma=6.5\ \mu Jy\ beam^{-1})}$ with contour levels [3, 5, 10, ..., 30, 35]$\times\sigma$.
The yellow contours indicate the ALMA Cycle 4 Band 6 continuum of the Companion component with $[3, 5]\times\sigma$ levels, where $\sigma=10.7\mathrm{\ \mu Jy\ beam^{-1}}$.
The white contours indicate the VLA BnA-configuration observation, as explained in panel (b).
The green contours indicate the VLA 8.2 GHz observation \citep{2000AAS..145..121P}.
(b) VLA BnA-configuration image $\mathrm{(\sigma=0.018\ mJy\ beam^{-1})}$. 
The white contours start at 4$\sigma$ and increase by a factor of 1.5 with a step of $3\sigma$.
The contour levels are given by $\mathrm{[start, start+step, start+(1+factor)*step, start+(1+factor+factor^2)*step, ...]}$.
(c) VLA B-configuration image $\mathrm{(\sigma=0.024\ mJy\ beam^{-1})}$. 
The white contours start at 5$\sigma$ and increase by a factor of 1.5.
(d) VLA A-configuration image $\mathrm{(\sigma=0.01\ mJy\ beam^{-1})}$.
The white contours start at 5$\sigma$ and increase by a factor of 1.5 with a step of $10\sigma$.
(NW) Zoom in on the NW galaxy. 
The white contours indicate the VLA BnA-configuration.
(SE) Zoom in on the SE galaxy.
The white and green contours indicate the VLA BnA- and A-configuration, respectively.
In all images, the black contours indicate the ALMA Cycle 6 Band 6 continuum, the yellow contours indicate the ALMA Cycle 4 Band 6 continuum imaging of the Companion component, and the yellow dashed line indicates the direction of the propagation of the radio jet.
The beam size of each false-color image is shown at the lower left by the white ellipse.
The new observations have identified a radio core that in general overlaps with the NW galaxy.
The Companion component lies 10 kpc southeast of the SE galaxy in projection.
The Companion and SE and NW galaxies are nearly in alignment with the propagation direction of the collimated radio jets.
The ALMA Cycle 6 observation reveals a sub-component, adjacent to the SE galaxy, that perfectly coincides with the SE hotspot identified by VLA 4.7 and 8.2 GHz observations.}
\label{fig:imaging}
\end{center}
\end{figure*}

\begin{table*}
\caption{Flux Measurements of the Dragonfly galaxy}
\label{tab:flux}
\begin{tabular}{ccccccccc}
\hline
Frequency & Component & $S_\nu$ & Spectral Index & $L_\mathrm{radio}$ & $P_{\nu_0}$ & Aperture & Ref.\\
(MHz) & & (mJy) & & (W Hz$^{-1}$) & $\mathrm{(erg\ s^{-1})}$ & & & \\
(1) & (2) & (3) & (4) & (5) & (6) & (7) & (8) \\
\hline
74 & Total & $3100\pm620$ & - & $(2.9\pm0.6)\times10^{28}{}^a$ & $(2.2\pm0.4)\times10^{43}$ & $80\arcsec$ & 1\\
147 & Total & $2620\pm10$ & $ -0.25\pm0.29$ & $(3.2\pm1.0)\times10^{28}$ & $(4.7\pm1.5)\times10^{43}$ & $25\arcsec$ & 2 \\
365 & Total & $1580\pm30$ & $ -0.56\pm0.02$ & $(2.7\pm0.1)\times10^{28}$ & $(9.9\pm0.3)\times10^{43}$ & $6\arcsec$ & 1 \\
$1.4\times10^3$ & Total & $453\pm91$& $ -0.93\pm0.15$ & $(1.2\pm0.3)\times10^{28}$ & $(1.6\pm0.4)\times10^{44}$ & $45\arcsec$ & 1 \\
$4.7\times10^3$ & Total & $115\pm12$ & $-1.13\pm0.19$ & $(3.6\pm0.8)\times10^{27}$ & $(1.7\pm0.4)\times10^{44}$ & - & 3\\
$8.2\times10^3$ & Total  & $47\pm4.7$ & $-1.61\pm0.25$ & $(2.5\pm0.6)\times10^{27}$ & $(2.0\pm0.4)\times10^{44}$ & - & 3 \\
$44\times10^3$ & Total & $2.42\pm0.32$ & $ -1.77\pm0.08$ & $(1.5\pm0.2)\times10^{26}$ & $(6.7\pm1.1)\times10^{43}$ & - & 4\\
$4.7\times10^3$ & NW & $7.2\pm0.72$ & - & - & - & $1\farcs16\times0\farcs44$ & 3\\
$8.2\times10^3$ & NW & $2.8\pm0.28$ & $-1.7\pm0.25$ & $1.6\times10^{26}$ & $1.3\times10^{43}$ & $0\farcs72\times0\farcs24$ & 3 \\
$44\times10^3$ & NW & $0.29\pm0.07$ & $-1.35\pm0.16$ & $(1.2\pm0.3)\times10^{25}$ & $(5.1\pm1.5)\times10^{42}$ & $0\farcs4\times0\farcs2$ & 4  \\
$4.7\times10^3$ & SE & $95.9\pm9.6$ & - & - & - & $1\farcs16\times0\farcs44$ & 3 \\
$8.2\times10^3$ & SE  & $44\pm4.4$ & $-1.4\pm0.25$ & $1.9\times10^{27}$ & $1.5\times10^{44}$ & $0\farcs72\times0\farcs24$ & 3  \\
$44\times10^3$ & SE & $1.88\pm0.09$ & $ -1.87\pm0.07$ & $(1.3\pm0.1)\times10^{26}$ & $(5.9\pm0.3)\times10^{43}$ & $0\farcs5\times0\farcs4$ & 4  \\
$237\times10^3$ & SE & $0.12\pm0.01$ & $-1.64\pm0.06$ & $(6.6\pm0.7)\times10^{24}$ & $(1.6\pm0.2)\times10^{43}$ & $0\farcs07\times0\farcs12$ & 4  \\
$44\times10^3$ & Core & $0.23\pm0.06$ & - & - & - & $0\farcs4\times0\farcs3$ & 4 \\
\hline
\end{tabular}
\begin{tablenotes}
{\item \raggedright Column (1): frequency.
Column (2): the measurement of the individual component.
Total$=$NW+SE+core.
See texts for details.
Column (3): Integrated flux densities over a certain region.
For flux measurements from references 1, 2, and 3 in Column (8), the radio emissions are unresolved such that the integration region can be considered as the beam size.
Column (4): the spectral index (positive convention, $S_\nu\propto\nu^\alpha$) between two observed frequencies, one in this line and the other in the line just above.
Column (5): the spectral luminosity $L_\nu=S_\mathrm{obs}4\pi D^2_L/(1+z)^{1+\alpha}$, where $S_\mathrm{obs}$ is the observed integrated flux densities and $D_L$ is the luminosity distance.
Column (6): the radio power $P_{\nu_0}=4\pi D^2_L(1+z)^{-\alpha-1}S_{\nu_0}\nu_0$ \citep{2010ApJ...720.1066C}.
Column (7): the aperture size used to integrate the flux density.
Column (8): references to flux measurements. 1: \cite{2010AA...511A..53V}; 2: \cite{2018MNRAS.474.5008D}; 3: \cite{2000AAS..145..121P}; 4: this work.\\
$^a$ The radio luminosity is calculated assuming the spectral index $\alpha=0$}
\end{tablenotes}
\end{table*}

\section{Results}
\subsection{Spatial Distribution}
In Fig.~\ref{fig:imaging} we show the imaging results of VLA 44 GHz and ALMA Band 6 observations, supplemented with previous VLA observation centred at 8.2 GHz \citep{2000AAS..145..121P}.
Observed by VLA at 8.2 GHz (green contours in Fig.~\ref{fig:imaging}), the radio emission shows clear double components that potentially correspond to the hotspots where the radio jet launched from the AGN interacts with the ambient medium.
The double hotspots are also visible in the VLA observation at 4.7 GHz \citep{2000AAS..145..121P}.
Thanks to the high flux density, the SE hotspot is clearly detected in all VLA 44 GHz observations, while the NW hotspot falls below a $5\sigma$ detection threshold in the B-configuration.
South of the SE hotspot, there is a diffuse emission region that may be corresponding to the expanding radio lobe, which is clearly shown in the B-configuration imaging.
We draw a yellow dashed line in Fig.~\ref{fig:imaging} to indicate the direction of the propagation of the radio jets under the assumption that the radio hotspots originate from the well-collimated jets.

In addition to the double hotspots, another radio emission is detected at a position very close to the NW galaxy in BnA-configuration (see panel (NW) in Fig.~\ref{fig:imaging}), while it remains non-detection in VLA B- and A-configurations.
Although slightly spatially offset by $\sim0\farcs06$, which is insignificant compared with the beam size $0\farcs14\times0\farcs08$, this radio source coincides well with the location of the NW galaxy in which the AGN resides.
This source also lies on the dashed line that links the double hotspots, suggesting that it can be the radio core -- the launch point of the bipolar radio jets -- of the RLAGN.

One serendipitous and interesting finding is a sub-component adjacent to the SE galaxy through the ALMA Band 6 observation (see panel (SE) in Fig.~\ref{fig:imaging}).
This sub-component has its location perfectly coincided with the SE hotspot, which is identified in VLA observations, that is supposed to be dominated by synchrotron radiation arising from the in situ acceleration due to a jet-ISM interaction.
However, the rest-frame frequency of the ALMA observation is $\sim700$ GHz which corresponds to $\sim400\ \micron$, entering the far-infrared (FIR) regime.
A mixture of different emission mechanisms, including \black{dust thermal emission, synchrotron radiation, and free-free emission}, can contribute to the observed flux density of this sub-component.
\black{
The $\mathrm{SFR}$ of the entire system is $\sim3000\ M_\odot\ yr^{-1}$, adopting a power-law black body radiation \citep{2014A&A...566A..53D}.
The integrated flux density of this sub-component is about 10 per cent of the dust thermal emission from SE and NW galaxies at 237 GHz.
Then, if this sub-component is dominated by jet-induced star formation, it may have an $\mathrm{\sim300\ M_\odot\ yr^{-1}}$, which corresponds to a synchrotron flux density of $3\ \mu$Jy and free-free flux density of $10\ \mu$Jy.
Therefore, synchrotron and free-free radiations attributed to the SFR contribute negligible contaminations to the sub-component at 237 GHz.
However, since a jet-induced star formation cannot be confirmed and its associated dust thermal emission cannot be evaluated in the current phase, we treat this sub-component as dominated by synchrotron radiation related to AGN activities throughout this work.
}

\subsection{Spectral Energy Distribution (SED)}
\begin{figure}
\begin{center}
\includegraphics[width=\columnwidth]{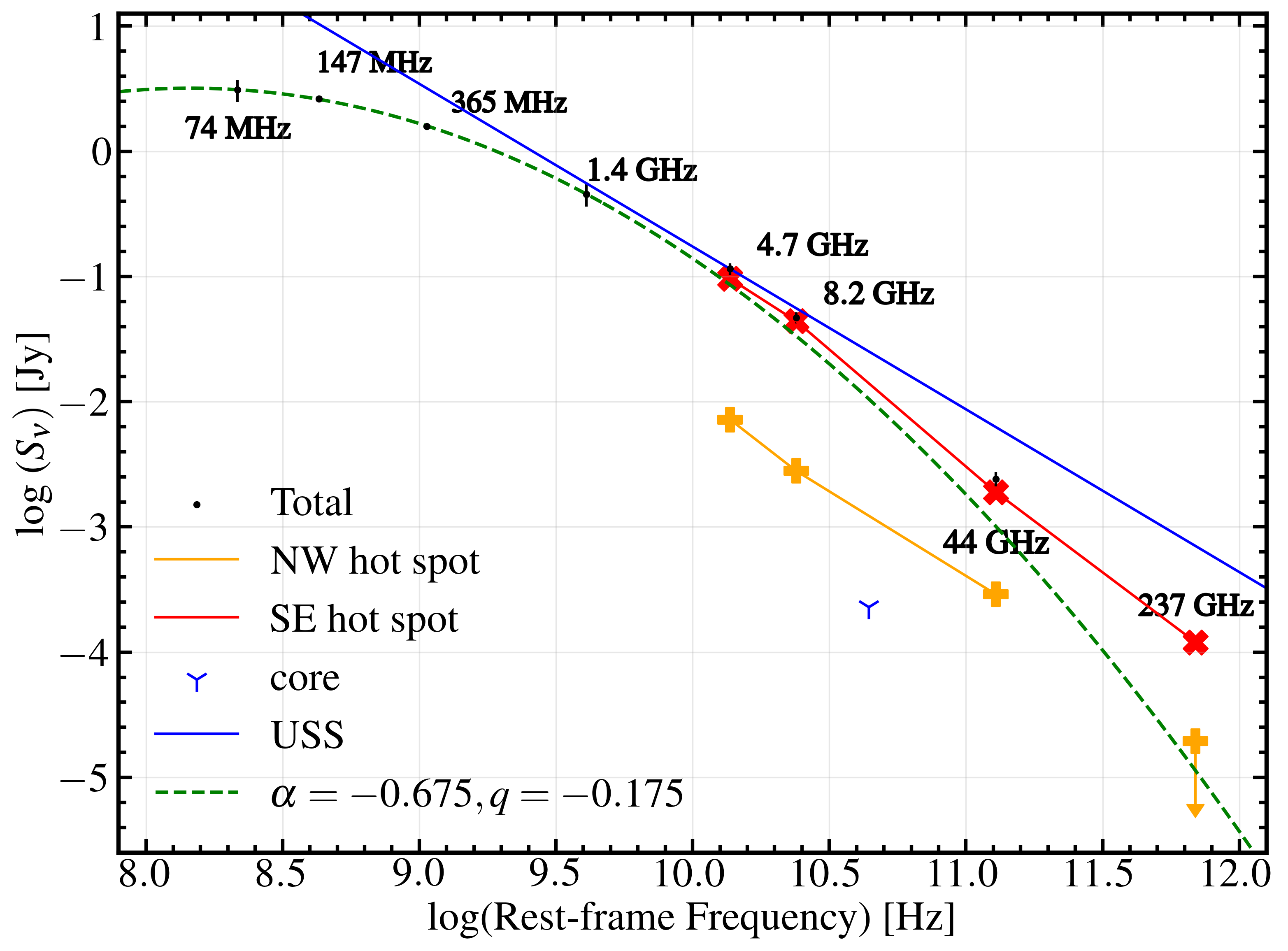}
\caption{Integrated flux density against frequency of the Dragonfly galaxy based on the measurements given in Table.~\ref{tab:flux}.
\black{The horizontal axis shows the rest-frame frequency and the observed frequency is shown by the data points.}
The green dashed line indicates the best-fitting of the observed $S_\mathrm{Total}$ by adopting a curved power-law model.
The NW hotspot is not detected at $3\sigma$-level in ALMA Cycle 6 and we set $3\sigma$ as an upper limit for its flux density to proceed with the discussion in \S\ref{sec:doppler boosting}.
The blue solid line indicates a spectral index $\alpha=-1.3$ and a spectrum steeper than this is classified as a USS source \citep{2000A&AS..143..303D}.
\black{Because of the smallness of uncertainties, some error bars are not clearly visible in the figure but can be found in Table~\ref{tab:flux}}.
\label{fig:sed}}
\end{center}
\end{figure}

Although the Dragonfly galaxy has been observed at low frequencies over decades, the low angular resolutions leave the radio emissions unresolved.
Hence, there is no available structural information on radio emissions in observations at $\tnr{\nu}{obs}\leq1.4$ GHz.
We listed the available archived data at different frequencies in Table~\ref{tab:flux} and plot the integrated flux density against frequency in Fig.~\ref{fig:sed}.
\black{The specific luminosity has exceeded $\mathrm{10^{28}\ W\ Hz^{-1}}$ at $\tnr{\nu}{obs}\leq1.4$ GHz, making this object an unambiguous radio-bright AGN \citep{2012ApJ...744...84Y,2016ApJ...831..168K,2020NewAR..8801539H}, though this radio luminosity can be smaller than that of the extreme populations found at $z>5$ by two orders of magnitude \citep{1999ApJ...518L..61V,2018MNRAS.480.2733S}.}

Based on the imaging results, we divide the radio emissions into three components: radio core, SE, and NW hotspots, and the total is the sum of these components.
All observations at $\leq1.4$ GHz have no resolved components, thus the integrated flux densities of these observations are simply treated as the sum of all three components.

As shown in Fig.~\ref{fig:sed}, the radio spectrum of the sum of different components shows no apparent turnover point, below which the spectral index becomes positive due to SSA and/or FFA and the medium is optically-thick to the radio emission \citep{2016era..book.....C}.
The globally negative spectral index suggests that the medium is likely to remain optically thin to the radio emission.
However, we do observe a gradual flattening of the slope towards lower frequencies, by which we cannot exclude the possibility that there can be a broad peak around 74 MHz at the observed frequency because of SSA/FFA.

At 4.7 and 8.2 GHz, we treated the peak flux densities of NW and SE hotspots as the integrated flux densities, i.e., $S_\mathrm{peak}=S_\mathrm{total}$, since both NW and SE hotspots are marginally resolved \citep{2001A&A...365..392P, 2008ApJ...681.1129B}.
At $\geq4.7$ GHz, the SE hotspot dominates over the observed flux densities, indicating that the SE hotspot may be Doppler boosted (see \S\ref{sec:doppler boosting}).
The slope of the SE hotspot becomes slightly shallower at 237 GHz than that at 44 GHz \black{$(\mathrm{\alpha=-1.87^{+0.07}_{-0.07}\to-1.64^{+0.06}_{-0.06}})$} because of a possible mixture of various emission mechanisms other than pure synchrotron radiation \black{at 237 GHz.
Otherwise, since contaminations from the radio lobes at 8.2 GHz may result in an overestimation of the spectral index at 44 GHz (see \S\ref{bimodal:contamination} and \S\ref{sec:caveat}), this slight flattening is not robust.}
The NW hotspot falls below the $3\sigma$ detection limit in ALMA Cycle 6 observation and has its slope steepened at 4.7 and 8.2 GHz, but becoming shallower at 44 GHz (see column (4) in Table.~\ref{tab:flux}).

Since the radio core is only observed at 44 GHz, we here estimate its contribution to flux densities at other frequencies and investigate its impact on the global shape of the total radio SED.
At 44 GHz, the radio core contributes to less than 10\% of the sum of different components, thus the absence of a core has negligible influence on the total radio SED.
As we will see in \S\ref{bimodal:contamination}, the radio core contributes to less than 1 per cent of the observed flux densities at 4.7 and 8.2 GHz.
Therefore, the contamination from the radio core cannot alter the shape of the observed spectral slope.

\section{The Curved SED}
\subsection{Compact-Steep-Source}
To better investigate the spectrum as a curve, we fit the observed flux densities at frequencies $\leq8.2$ GHz by a curved power-law model given by \citep{2017ApJ...836..174C}
\begin{equation}
    S_\nu=a\nu^\alpha e^{q(\ln\nu)^2},
\end{equation}
where $S_\nu$ is the flux density at the observed frequency $\nu$ in GHz, $\alpha$ is the spectral index, $a$ is a factor that determines the amplitude of the non-thermal spectrum, and $q$ is the curvature parameter that optimizes the spectral curvature, by which the peaked frequency (or turnover frequency) is defined as $\nu_\mathrm{peak}=e^{-\alpha/2q}$.
The best-fitting result of the sum of integrated flux densities is shown in Fig.~\ref{fig:sed} by the green dashed line with $q=-0.175\pm0.01$ and $\alpha=-1.05\pm0.03$, showing no evidence of a significant curvature ($|q|\geq0.2$, \citealt{2012MNRAS.421..108D}).
The corresponding peaked frequency is $\nu_\mathrm{peak}=e^{-\alpha/2q}\sim50-60$ MHz.
The curved power-law colossally deviates from the observational data higher than 44 GHz due to the energy loss, which will be studied in \S\ref{ci}.

Since the projected linear separation of the two hotspots is only $\sim1\farcs5$ ($\sim13$ kpc at $z=1.92$), the Dragonfly galaxy can be classified as a CSS source \citep{2021A&ARv..29....3O}.
By fitting a single Gaussian component to the image plane, the size of the SE hotspot deconvolved from the beam is $(0\farcs054\pm0\farcs015)\times(0\farcs033\pm0\farcs01)$, corresponding to a radius of $r\sim~0.47$ kpc at $z=1.92$, which yields a typical hotspot size of CSS sources \citep{2000MNRAS.311..397J}, providing additional support for a CSS classification.
Using an empirical relation that correlates the projected angular size ($l$ in kpc) and $\nu_\mathrm{peak}$ (in GHz) for CSS sources \citep{1997AJ....113..148O},
\begin{equation}
    \log \nu_\mathrm{peak}=-0.21(\pm0.05)-0.65(\pm0.05)\times\log l,
\end{equation}
the estimated turnover frequency is $116\pm20$ MHz for $l\sim13$ kpc.
This estimation is close to the value given by the curved power-law but can be underestimated since the observation may not detect the full extent of the synchrotron radiation due to the diffuse emission attributed to radio lobes (see panel (c) in Fig.~\ref{fig:imaging} and the discussion in \S\ref{bimodal:contamination}).

CSS sources can be young AGNs with evolving radio jets or can be old populations that have their radio jets confined by the dense ISM of their host galaxies \citep{2021A&ARv..29....3O}.
A young CSS source may merely have intermittent or transient activities as well, incapable of forming large structures \citep{2021A&ARv..29....3O}.
However, confined by the dense ISM of the AGN host galaxy, the radio jet can strongly couple with the surrounding material, leading the jet to slow down and lose kinetic energy.
When the jet is not sufficiently powerful to overcome the confining pressure of the ISM, such strong couplings may result in the disruption of the jet and the formation of radio lobes \citep{2016MNRAS.461..967M,2023arXiv230111937B}.
In the Dragonfly, since the jet has obviously left its host galaxy, i.e., the NW galaxy, and interacted with the ambient medium to create the hotspots, confinement by the host galaxy is not favored.
Therefore, the RLAGN residing in the Dragonfly galaxy is likely to be a CSS source that is associated with an AGN at an early evolution stage \citep{2012ApJ...745..172D}.
A CSS source that launches powerful radio jets can escape more quickly from the confining medium than their low-power counterparts and may evolve into a Fanaroff-Riley II (FRII) radio galaxy \citep{1998PASP..110..493O,2016MNRAS.461..967M,2017ApJ...836..174C}.
Nonetheless, we have to consider the possibilities of transient or intermittent radio AGN activities and this requires a further investigation into the synchrotron age.

\subsection{Synchrotron Ageing}\label{ci}
At higher frequencies, there is no cut-off but a steepening of the slope at $\nu_\mathrm{obs}\geq4.7$ GHz with $\alpha<-1.3$.
Such a steep slope in HzRGs is often found at $\sim1.4$ GHz in the radio SED of USS sources \citep{2000A&AS..143..303D}.
One possible interpretation for this feature is that the increasing cosmic microwave background (CMB) radiation at higher redshifts causes stronger inverse-Compton (IC) losses, and the strength of the magnetic field equivalent to the CMB $(B_\mathrm{CMB})$ dominates over the magnetic field \citep{2017MNRAS.469.4083S,2019SSRv..215...16V}.
However, in the SE hotspot, adopting a lower limit of the magnetic field in energy equipartition $B\mathrm{_{eq}}$, \citet{2000AAS..145..121P} gives $B\mathrm{_{eq}}=160\ \mu$G, which is much higher than the expected $B_\mathrm{CMB}$ (see \S\ref{sec:fitting}).
Therefore, for the Dragonfly galaxy, the dominated energy loss at high frequency should be explained by synchrotron ageing, that is, since the electron energy loss rate $-dE/dt$ is proportional to $E^2$, higher-energy electrons deplete their energy faster.

\begin{figure}
\begin{center}
\includegraphics[width=\columnwidth]{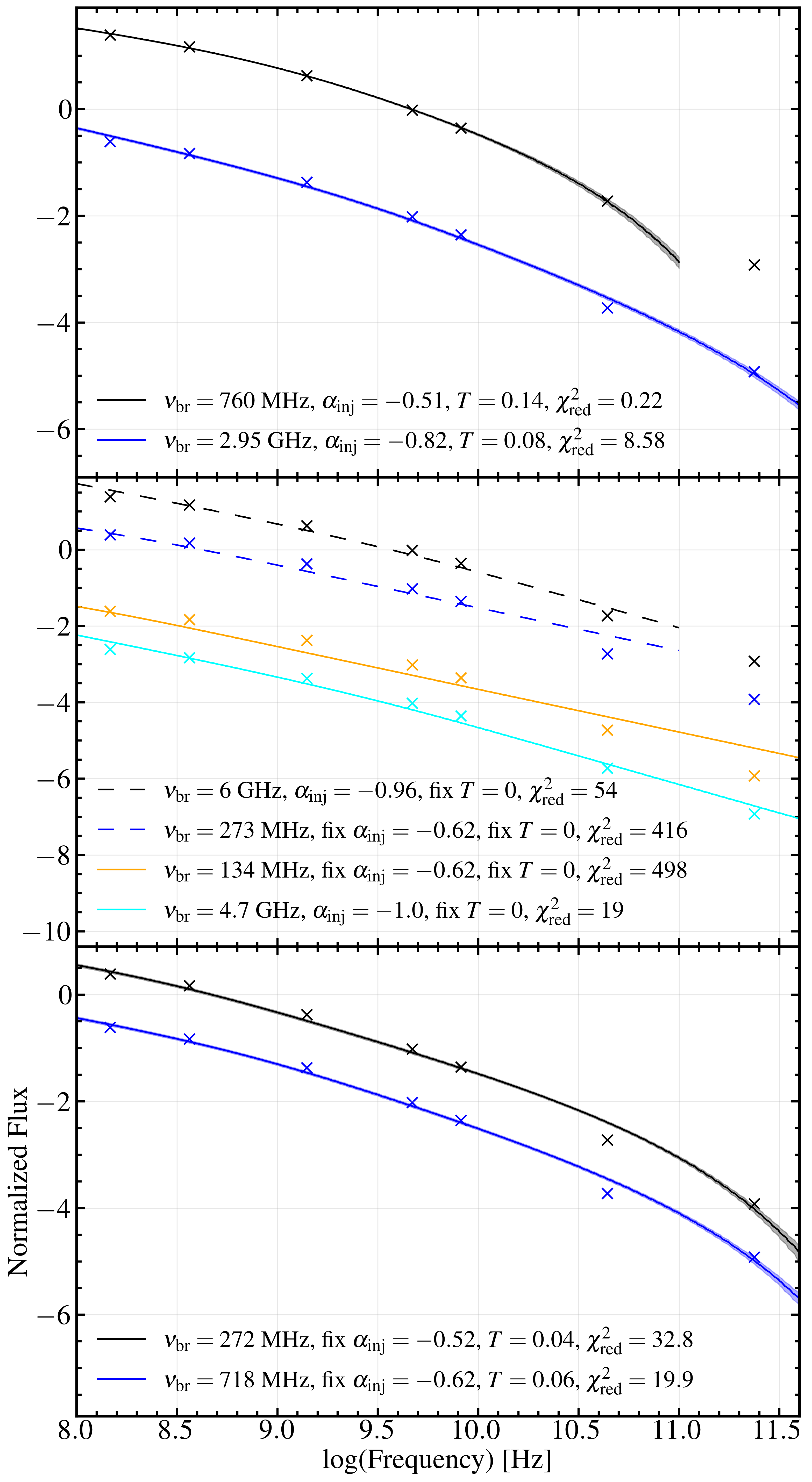}
\caption{Fitting results of the SE hotspot using \textsc{synchrofit}.
The vertical axis is in an arbitrary unit and all data points are the same but shifted along the vertical axis for display purposes.
The shaded regions embed 2$\sigma$ of the modeled spectrum and $\chi_\mathrm{red}^2$ indicates the reduced-$\chi^2$ statistics.
\textit{Top}: The black and blue lines indicate the fittings without and with ALMA data, respectively.
\textit{Middle}: Fittings with the remnant fraction $T$ fixed to 0.
The dashed and solid lines indicate the fittings without and with ALMA data, respectively.
Without a remnant fraction $T$, the fitting is almost a straight line and cannot reproduce the curvature of the observed SED.
\textit{Bottom}: Fittings with ALMA data and fixed injection spectral index $\alpha_\mathrm{inj}$.
The $\alpha_\mathrm{inj}$ is chosen to lie within the range of a DSA scenario.}
\label{fig:synchrofit}
\end{center}
\end{figure}

\subsubsection{Synchrotron Fitting}\label{sec:fitting}
The injection of new synchrotron-emitting electrons can be described by a power-law spectrum that is referred to as the injection spectrum -- a well-established method to investigate high-frequency energy losses.
The injection spectrum is the ensemble of all relativistic electron populations with an injection spectral index $\alpha_\mathrm{inj}$ that describes the slope of a synchrotron radiation spectrum.
The CI (continuous injection) model describes a source with continuous replenishment of new relativistic particles over its lifetime \citep{1970ranp.book.....P}.
The JP (Jaffe-Perola) and KP (Kardashev-Pacholczyk) models describe a source that undergoes a single burst of particle acceleration and then ages rapidly, with and without a continuous isotropization of the electron pitch angle, respectively \citep{2019SSRv..215...16V,1973A&A....26..423J,1970ranp.book.....P,1962SvA.....6..317K}.

We fit the radio SED of NW and SE hotspots using \textsc{synchrofit} to estimate the break frequency $\nu_\mathrm{br}$, above which the synchrotron radiation ages, and injection index $s$ that reflects the initial power-law of the electron energy distribution \citep{2018MNRAS.474.3361T,2018MNRAS.476.2522T}.
The injection spectral index can be calculated from $\alpha_\mathrm{inj}=(1-s)/2$.
The fitting also returns an estimate of the remnant fraction $T$, which is defined as $T=t_\mathrm{off}/\tau$, where $t_\mathrm{off}$ is the time the source spent in an inactive phase and $\tau=t_\mathrm{off}+t_\mathrm{s}$ is the total source age, where $t_\mathrm{s}$, synchrotron age, is the duration of the continuous injection.
The synchrotron age is determined by $\nu_\mathrm{br}$ together with the magnetic field strength $B$.
The uncertainties are estimated using 1000 Monte Carlo iterations.
The number of increments used to sample the allowing ranges for free parameters $\nu_\mathrm{br}$, $s$, and $T$, is set by default.
Three iterations have been performed such that the free parameters fitted by the former one of two consecutive iterations will be fed to the latter one as initial guesses.

Prior to performing the fitting, we assume that the integrated flux densities in unresolved observations ($\tnr{\nu}{obs}\leq$1.4 GHz) only emanate from the NW and SE hotspots, with negligible contamination from NW and SE galaxies and the radio core for the reasons stated in \S\ref{bimodal:contamination}.
To divide the flux densities of unresolved components into those of SE and NW hotspots, we adopt a flux ratio that follows the values found in 4.7 and 8.2 GHz observations, that is $S_\mathrm{SE}:S_\mathrm{NW}=15:1$.
Further fittings considering different flux ratios and contaminations from the radio lobes are presented in Appendix \ref{app:synchrofit}.
We only present the fitting results using observational data at $\geq147$ MHz here since a shallower slope towards 74 MHz due to ionization losses may aggravate the fitting results, though an inclusion or omission of 74 MHz data does not affect our conclusion (see Appendix \ref{app:synchrofit}).

Previous studies have shown that the injection spectrum of a radio hotspot is better described by the CI model, which is natural considering the radio jet continuously accelerates the electrons in situ \citep{1991ApJ...383..554C,2011A&A...526A.148M,2020A&A...634A...9M}.
Additionally, the reduced-$\chi^2$ of the JP and KP models is larger than that of the CI model by at least a factor of three.
Furthermore, we have performed fitting by fixing $T=0$, assuming that there is no quiescent phase during the jet-medium interaction.
However, as clearly shown in Fig.~\ref{fig:synchrofit}, with or without ALMA 237 GHz data, the observed curvature of the SED of the SE hotspot cannot be reproduced by the algorithm in $T=0$ cases, and this scenario can be rejected.
Hence, we only discuss the fitting results of the CI model with non-zero $T$ here.

\subsubsection{Fitting Results}\label{sec:synchro result}
The synchrotron age, i.e., the time-scale during which the continuous acceleration lasts, is calculated by \citep{2018MNRAS.474.3361T,2020A&A...634A...9M}:
\begin{equation}
    \tnr{t}{s}=1610\frac{B_\mathrm{eq}^{1/2}}{B_\mathrm{eq}^2+B_\mathrm{CMB}^2}\frac{1}{\nu_\mathrm{br}(1+z)^{1/2}},
\end{equation}
where $B_\mathrm{CMB}=3.25(1+z)^2\ \mu$G is the magnitude of the magnetic field equivalent to the CMB.
The estimation of $B_\text{eq}$ in an energy equipartition condition can be calculated using \citep{1980ARA&A..18..165M,2020A&A...634A...9M}:
\begin{equation}
    B_\mathrm{eq}=9.6\times10^{-12}\left(\frac{P_\text{1.4GHz}(1+k)}{V^3}\right)^{2/7},
\end{equation}
where $P_\text{1.4GHz}$ is the radio power at 1.4 GHz in a unit of W Hz$^{-1}$, \textit{k} is the protons to electrons energy ratio in the hotspot and set as 1, and \textit{V} is the hotspot volume in kpc$^3$. 
We corrected the observed flux density to its rest-frame considering the Doppler boosting effect (see \S\ref{sec:doppler boosting} and \S\ref{jet-ism}) to calculate $P_\text{1.4GHz}$.
The hotspot is assumed to occupy an elliptical cylinder volume, in which the cross-section corresponds to the beam size and the height is the transverse size of the hotspot.
The estimated $B_\text{eq}$, taking a Doppler boosting factor $\delta=2-4$ (see the discussion in \S\ref{jet-ism}), is $175-213\ \mu$G for the SE and $214-261\ \mu$G for the NW hotspot, respectively.
These estimates lie within $B\mathrm{_{eq}=160-700\ \mu G}$ given by \citet{2000AAS..145..121P} for a sample of HzRGs including the Dragonfly galaxy.

The NW hotspot is fitted without 237 GHz data, and the fitting result is $\nu_\mathrm{br}=980\pm180$ MHz, $\alpha_\mathrm{inj}=-0.59\pm0.01$, and $T=0.08\pm0.02$.
Considering a quiescent phase during the continuous acceleration, the above fitting gives  $t_\mathrm{s}\sim2-3\times10^5$ yr.

Without ALMA 237 GHz data, the SE hotspot has $\nu_\mathrm{br}=760\pm140$ MHz, $\alpha_\mathrm{inj}=-0.51\pm0.01$, $T=0.13\pm0.03$, and $t_\mathrm{s}\sim3-4\times10^5$ yr.
With ALMA 237 GHz data, i.e., under the assumption that the sub-component observed at 237 GHz has synchrotron-dominance, the SE hotspot then has $\nu_\mathrm{br}=2.95\pm0.5$ GHz, $\alpha_\mathrm{inj}=-0.82\pm0.02$, $T=0.08\pm0.02$, and $t_\mathrm{s}\sim2\times10^5$ yr.
Although the model flux density well reproduces the observed value at 237 GHz in this case, it slightly exceeds the observation at 44 GHz (blue curve in the top panel of Fig.~\ref{fig:synchrofit}).
Naturally, one will expect a further steepening of the spectrum slope due to both synchrotron ageing and IC loss, and thus, at 237 GHz, the synchrotron radiation is supposed to have a much lower flux density than the currently observed value.
Considering that the 237 GHz data lies within the FIR regime at the rest-frame, a reasonable explanation is that the sub-component perfectly coincided with the SE hotspot and observed in ALMA Cycle 6 may be contaminated from thermal bremsstrahlung and/or the Rayleigh-Jeans tail of the thermal dust emission.
Another explanation is that, as we will see in \S\ref{bimodal:contamination} and \S\ref{sec:caveat}, this may be a result that the undetected diffuse radio lobes have contaminated low-resolution observations such that the observed flux densities of the SE hotspot at $\leq8.2$ GHz used for the fitting are over-assumed.
Accordingly, at 44 GHz, the true spectral slope should be shallower than the current one and steepens towards 237 GHz.

The First-order Fermi Acceleration, or diffusive shock acceleration (DSA), is one important mechanism for particle acceleration due to jet-medium interaction \citep{1991ApJ...383..554C,2019SSRv..215...16V}.
In a non-relativistic shock acceleration case, the injection index can be expressed by the compression ratio $\chi$ of downstream to upstream proper particle densities \citep{2009herb.book.....D}.
For a strong shock with velocity $u\gg c_s$, where $c_s$ is the sound speed, in a uniform ISM, $\chi=4$, corresponding to $\alpha_\mathrm{inj}=-0.5$.
For the SE hotspot, without 237 GHz data, the calculated $\alpha_\mathrm{inj}=-0.51$ agrees well with this scenario.
While for $\alpha_\mathrm{inj}=-0.59$ found in the NW hotspot, the shock dynamics may be altered by the relativistic particles slightly \citep{1991ApJ...383..554C}.
When the ALMA 237 GHz data is considered, the SE hotspot has $\alpha_\mathrm{inj}=-0.82$, which is significantly different from the values expected for the case above.
Based on current observations, we cannot decompose the 237 GHz data into different emission mechanisms which may contaminate the observed flux density of pure synchrotron radiation.
As a result, the modeled flux density at 237 GHz and its corresponding $\alpha_\mathrm{inj}=-0.82$ can be an overestimation.
We further fitted the spectrum with fixed $\alpha_\mathrm{inj}$ and found that as $\alpha_\mathrm{inj}$ increases, $T$ and  $\nu_\mathrm{br}$ decrease (see Fig.~\ref{fig:synchrofit}).
Therefore, albeit there can be a mixture of different emission mechanisms, in this case, we treat the current fitting up to 237 GHz data with $\alpha_\mathrm{inj}=-0.82$ as a lower limit for the synchrotron age.
Moreover, if $\alpha_\mathrm{inj}=-0.82$ is robust, it may reflect significant environmental differences between SE and NW hotspots.
The SE hotspot might reside in a denser environment such that the fluid velocity of upstream relativistic particles $\tnr{v}{up}$ becomes smaller, resulting in a decrease in $s$ which is given by $s=\frac{2+\chi}{\chi-1}=\frac{v_\mathrm{up}+2v_\mathrm{down}}{v_\mathrm{up}-v_\mathrm{down}}$, where $\tnr{v}{down}$ is the downstream fluid velocity \citep{2009herb.book.....D}.
Accordingly, the injection spectral index steepens.

The synchrotron ages of both NW and SE hotspots have an order of $10^5$ yr, lying within the order of magnitude $10^3-10^5$ yr observed in typical CSS sources that are argued to be related to AGNs at an early evolution stage \citep{1998PASP..110..493O,2003PASA...20...19M}.
Since a remnant fraction is essential to fit the observed curved SED by a CI model, the RLAGN hosted by the Dragonfly galaxy is not only young but may also have intermittent or transient activities.
This may indicate that the central SMBH is in a transition phase towards an active one because the gas inflow arising from the merging events fuels the SMBH growth.

\section{Discussions}
In this section, we first discuss the Doppler boosting effect because of the highly imbalanced flux densities between the SE and NW hotspots.
However, the flux ratio of SE to NW changes from $S_\text{SE}/S_\text{NW}\approx15:1$ at 4.7 and 8.2 GHz to $S_\text{SE}/S_\text{NW}\approx6:1$ at 44 GHz, suggesting that there is contamination in low-resolution observations and/or the intrinsic flux densities of the SE and NW hotspots may differ.
We then discuss the possible scenarios.
Next, considering the proximity of the sub-component to the SE galaxy, we discuss the energetic input of the radio jet into the ISM.
Finally, we discuss the caveats to the results and conclusions we make in this work.

\subsection{Doppler Boosting?}\label{sec:doppler boosting}
The radio core and the SE and NW hotspots are well aligned in a straight line.
However, the flux densities between SE and NW hotspots are highly imbalanced.
This feature may be a result of the significantly different environments between SE and NW hot spots, which is an implication of the fitting results of the SE hot spot including ALMA 237 GHz data.
If this is the case, adopting equation (9) in \citet{2019EPJWC.21004006A}, the upstream energy density of relativistic electrons of the SE hotspot can be larger than that of the NW one by an order of magnitude.
However, an investigation into the detailed environments is impossible with the current data set.

Here we consider the Doppler boosting effect as another scenario to explain the flux density imbalance.
In this case, the SE hotspot is associated with the approaching jet while the NW one, as the counterpart, is associated with the receding jet.
Therefore, the SE hotspot has its observed flux density Doppler boosted arising from relativistic beaming, and the NW hotspot is dimmed accordingly \citep{2016era..book.....C}.
On the other hand, the global structure is not perfectly symmetric since the projected distance between the hotspot and radio core is $l\mathrm{_{SE}=0\farcs472\sim4\ kpc}$ for SE and $l\mathrm{_{NW}=0\farcs678\sim5.8\ kpc}$ for NW hotspot.
This can be explained by a misalignment angle between the two jets because the initially inclined jet may interact with the gas in the circumnuclear disk, resulting in a further bending of the jet \citep{2018MNRAS.479.5544M,2019A&A...632A..61G,2022MNRAS.514.4535T}. 
This can also be explained by that the SE jet has encountered the ISM of the SE galaxy, resulting in a shorter jet-traveled distance compared with the NW jet that may interact with the intergalactic medium.

The enhancement or dimming of the intrinsic flux density is dependent on the spectral index and Doppler factor $\delta$ (see discussion in \S\ref{jet-ism}) which is determined by the inclination angle and advance speed of the jet head.
With observations of NW and SE hotspots at multiple frequencies, the flux density ratios of the approaching and receding one provide constraints on the parameter space of jet geometry.
Denote the flux density ratio by $R$, one can expect \citep{2019A&A...631A..49K,2012ApJS..198....5A}
\begin{equation}
    R=\frac{S_\mathrm{adv}}{S_\mathrm{rec}}=\frac{S_\mathrm{SE}}{S_\mathrm{NW}}=\left(\frac{1+\beta\cos{\theta}}{1-\beta\cos{\theta}}\right)^{3-\alpha},
\end{equation}
where $\theta$ is the inclination angle of the approaching jet relative to LOS, $\beta=v/c$ is the advance speed of jet head in the unit of light speed, and $\alpha$ is the spectral index, defined as $S_\nu\propto\nu^\alpha$.

Since there are two values of ratios dependent on the frequency, that is, $R\sim15$ at $\nu_\mathrm{obs}=4.7$ and 8.2 GHz and $R\sim6$ at $\nu_\mathrm{obs}=44$ and 237 GHz
\footnote{The values of $\alpha$ for each hotspot at 8.2, 44, and 237 GHz follow those listed in Table.~\ref{tab:flux}, and $\alpha=-1.13$ is assumed for both hotspots observed at 4.7 GHz.
The NW hotspot is not detected in ALMA Band 6.
We assumed an upper limit of the NW component based on $3\sigma$ level to estimate the flux ratio at 237 GHz, which gives $S\mathrm{_{NW,237GHz}=0.0194\ mJy}$.}, we find two ranges for $\beta$: $\sim0.2$c$-0.3$c for $R\sim15$ and $\sim0.1$c$-0.2$c for $R\sim6$. 
Both ranges are in agreement with powerful FRII radio galaxies that have $\beta$ varying within 0.1c -- 0.5c with mildly relativistic jets and giant radio lobes \citep{2008ApJ...681.1129B,2009A&A...494..471O}.
Therefore, a mildly relativistic scenario is valid regardless of the exact value of the flux density ratio.

\subsection{Bi-modal Flux Ratios}
\label{bimodal:contamination}
Since the low-resolution VLA 4.7 and 8.2 GHz observations have much larger flux ratios $(S_\mathrm{SE}/S_\mathrm{NW}\sim15:1)$ compared with high-resolution and high-frequency VLA 44 GHz observations $(S_\mathrm{SE}/S_\mathrm{NW}\sim6:1)$, we consider several possibilities to explain this difference.

First, the low-frequency radio emission of the SE hotspot is contaminated by the \black{synchrotron radiation, which is attributed to the cosmic-rays accelerated in the supernova remnants, and free-free emission, which originates from H~{\sc ii} regions}, from the SE galaxy located at $\sim1$ kpc away.
Although the Dragonfly galaxy is a starburst galaxy, its observed radio luminosities at 4.7 and 8.2 GHz are well beyond the star formation-powered synchrotron regime \citep{2016ApJ...831..168K}.
Following \citet{1992ARA&A..30..575C}, we estimated the \black{thermal and non-thermal flux densities associated with star formation at 8.2 GHz, with a spectral index $\tnr{\alpha}{non-thermal}=-0.75$, $\tnr{\alpha}{thermal}=-0.1$, and $\mathrm{SFR\sim3000\ M_\odot yr^{-1}}$ (Paper II; Zhong et al. in prep.).}
The resulting \black{$S_{\nu,\text{non-thermal}}\sim0.07$ and $S_{\nu,\text{thermal}}\sim0.11$ mJy are smaller than the observed flux density by more than two orders of magnitude}.
Hence this scenario can be undoubtedly rejected.

Radio core is the second possibility, but can be immediately rejected because the core component in general shows flatter spectral indices relative to radio hotspots and lobes \citep{1994AJ....108.1163K,2019A&A...630A..83Z}.
Assuming $\alpha_\mathrm{core}=-0.5$, the observed flux density $0.23$ mJy at 44 GHz corresponds to $\sim0.5$ mJy at 8.2 GHz and $\sim0.7$ mJy at 4.7 GHz.
This explains why the core is only identified in high-sensitivity VLA 44 GHz observation but remains non-detection in VLA 4.7 GHz and 8.2 GHz observations.

The third possibility is reflected in the spectral index between 8.2 and 44 GHz.
By adopting $\alpha_\mathrm{NW}=-1.35$ instead of $\alpha_\mathrm{SE}=-1.87$, the expected flux density of the SE hotspot at 44 GHz is $\sim4.5$ mJy, and the corresponding flux ratio becomes $S_\mathrm{SE}/S_\mathrm{NW}\approx15$.
Hence, it is clear that the steeper slope of the SE hotspot results in a change in flux ratios.
A plausible explanation for this steepening is that, in VLA 4.7 and 8.2 GHz observations, the unresolved SE component contains flux densities not only from the SE hotspot but also the expanding radio lobes associated with it.
At 44 GHz, the diffuse radio lobes are not fully imaged, leading to a significant decrease in the observed flux densities linked to the SE hotspot.
This is supported by the low-resolution Australia Telescope Compact Array continuum observation at 40 GHz that has a total flux density $S_\nu=5.1\pm1.5$ mJy \citep[Lebowitz et al. 2023;][]{2011ApJ...734L..25E}, which is about twice the sum of flux densities of the radio core and SE and NW hotspots observed at VLA 44 GHz.

Last but not least, the possibility that the intrinsic flux densities of SE and NW hotspots may differ exists.
In this scenario, this more rapid steepening can be explained by that the SE hotspot has lost more energy at higher frequencies compared with the NW one.
As discussed in \S\ref{sec:synchro result}, this might be a consequence that the SE jet has interacted with a medium with higher density than the NW one, resulting in a steeper initial power-law energy distribution of the injected electrons.
An additional explanation is that, due to the 3D geometry of the bipolar jets, the jet-traveled distance has a projection onto the LOS.
Then, the total projection will be $l_\text{LOS,total}=l_\text{LOS,NW}+l_\text{LOS,SE}$. 
This is an additional light-traveled distance for the synchrotron radiation from the NW hotspot.
Accordingly, the NW hotspot is intrinsically younger than the SE one by $l_\text{LOS,total}/$c yr at the time of observation, and thus has a higher magnetic field strength than the SE hotspot does.

\subsection{The Possible Jet-ISM Interaction}\label{jet-ism}
A serendipitous discovery of ALMA Band 6 continuum imaging is the sub-component adjacent to the SE galaxy and coincides with the location of the SE hotspot.
The radio continuum to the east of the SE galaxy is consistent with the distribution of CO(6-5) line emission regions.
However, west to the SE galaxy, no line emission has been detected around the SE hotspot and only the continuum is detected as the sub-component in high-resolution observations, while extended CO(6-5) line emission is found at this location in the low-resolution observation (Lebowitz et al. 2023; Paper II; Zhong et al. in prep.).
Additionally, there is only a faint rest-frame UV continuum at this location.
Furthermore, \citet{2015A&A...584A..99E} has found a giant molecular cloud lying close to the propagation direction of the SE jet, offset from the jet propagation axis by $\sim0\farcs3$ and indicated by the Companion in panel (a) in Fig.~\ref{fig:imaging}.
Considering also the small offset ($\sim1$ kpc) of the sub-component to the centroid of SE galaxy, this sub-component may indicate an interaction between jet and ISM, through which the radio jet has driven a massive molecular outflow.

To examine this scenario, we estimate the spatial separation of SE and NW galaxies, which is $\sim7$ kpc in the projected plane.
Assuming this Companion in the form of molecular gas that may reach an outflow velocity of $\sim1000$ km s$^{-1}$ \citep{2012ApJ...757..136W}, the Companion takes $\sim7$ Myr to travel such a distance. 
However, the upper limit of the lifetime of the SE hotspot has an order of $10^5$ yr, suggesting that the bulk of the massive molecular gas cannot be the outflow driven by the jet-ISM interaction.

Nonetheless, we can still estimate the intrinsic flux density of the SE hotspot through $\delta^{2+\alpha}<S/S_0<\delta^{3+\alpha}$ \citep{2016era..book.....C}, where $S$ is the observed flux density, $S_0$ is the intrinsic flux density,  $\delta\equiv[\gamma(1-\beta\cos\theta)]^{-1}$ is the Doppler factor and $\gamma\equiv(1-\beta^2)^{-1/2}$ is the Lorentz factor.
Using equation (4), we find the observed flux amplified by $\delta\sim2$ and $\delta\sim3-4$ for $S_\mathrm{SE}/S_\mathrm{NW}\sim6$ and $S_\mathrm{SE}/S_\mathrm{NW}\sim15$, respectively.
Conservatively, at $\mathrm{\nu_{obs}}\sim1.4$ GHz, we assume that SE hotspot contributes to $\sim90\%$ of the total observed flux densities and Doppler boosted by a factor of 2 as an upper limit, the corresponding intrinsic radio power is $P_\text{1.4GHz}=(7.3\pm1.9)\times10^{43}\ \mathrm{erg\ s^{-1}}$.

Adopting an empirical correlation between the radio power at 1.4 GHz and the kinetic power of the jet $P_\mathrm{jet}$ \citep{2010ApJ...720.1066C}, the approaching jet associated with the SE hotspot has a total jet kinetic power $P\mathrm{_{jet}=(7\pm9)\times10^{46}\ erg\ s^{-1}}$.
The total energy injection for the ambient gas to be accelerated by the radio jet can be estimated via an energy conservation argument \citep{2006ApJ...650..693N}:
\begin{equation}
    M_\mathrm{out}=2\times10^{11}E_\mathrm{out,60}v^{-2}_\mathrm{esc,500}\ \mathrm{M_\odot},
\end{equation}
where $E_\mathrm{out,60}$ is the energy of the outflow in units of $10^{60}$ ergs and $v^{-2}_\mathrm{esc,500}$ is the outflow velocity in units of 500 $\mathrm{km\ s^{-1}}$.
Even taking a lower limit on the SE jet lifetime, the total energy injection is $E\sim1\times10^{59}$ erg, suggesting that the jet kinetic energy can result in a significant molecular gas outflow of $M_\mathrm{out}\sim7\times10^9 \ \mathrm{M_\odot}$ even reaching an outflow velocity $\mathrm{\sim1000\ km\ s^{-1}}$ for the most powerful jets \citep{2012ApJ...757..136W}.
This may explain why there is merely diffuse CO(6-5) line emission in the west of the SE galaxy, that is, the molecular gas has been blown away by the jet-ISM interaction through the jet kinetic power, and the gas has been excited to higher excitation levels through the jet thermal power.

The jet-driven large-scale molecular outflows ($\tnr{M}{H_2}\sim10^{9-10}\ \mathrm{M_\odot}$), which indicates a removal of a significant fraction of the ISM from the AGN host galaxy, have been observed in some massive galaxies ($M_\star\sim10^{11}\ \mathrm{M_\odot}$) at high redshifts \citep[e.g.,][]{2006ApJ...650..693N,2007A&A...475..145N,2017A&A...600A.121N,2023arXiv230317484E}.
However, a jet-ISM interaction in the Dragonfly galaxy discussed here happens between the jet and the merging pair (SE galaxy) of its host galaxy (NW galaxy), making it a unique sample at high redshifts without parallels.
A detailed investigation into the AGN feedback and its associated outflow traced by CO(6-5) line emission through an interaction of the jet and the ISM of the NW galaxy will be presented in Paper II.

\subsection{Caveats}\label{sec:caveat}
The interpretation of the remnant fraction should be careful.
It may simply indicate that the jet has been switched off and the jet-medium interaction has ceased at the observation timing, which is a general understanding of non-zero remnant fraction in the CI model \citep{2011A&A...526A.148M,2018MNRAS.476.2522T,2020A&A...634A...9M}.
We further speculate on the interpretation of the remnant fraction in our work considering the complexity of the particle acceleration in realistic models.
The radio source may have experienced a quiescent time-scale after the first phase of continuous acceleration, following a second phase in which the restarted radio jets interact with the medium again and replenish a new population of relativistic electrons.
In this restarting scenario, if the restarted jet reaccelerates the particles universally, then the entire radio spectrum will be altered and the break frequency no longer traces the time-scale since the initial acceleration begins \citep{1991ApJ...383..554C}.
In this case, a flattening in radio SED at high frequencies will be observed because of the accelerated electrons.
However, this feature is not found in our target.
Then, the restarted jet could happen shortly before the observation timing or it is not sufficiently powerful to alter the ensemble of synchrotron-emitting electrons.
Therefore, compared with those dying radio sources, which have been inactive for a long period with $\tnr{t}{s}\sim10^6-10^7$ yr, $\tnr{B}{eq}\sim\text{tens}\ \mu$G, and remnant fraction of $T\geq0.2$, the small remnant fraction found in the Dragonfly galaxy indicates that this CSS source may have experienced a past radio activity, and our conclusion that the RLAGN may have transient or intermittent activities remains unchanged.

We note that the age estimations of the radio hotspots can be uncertain because of the magnetic field strengths, ambient medium densities, and volumes of hotspots.
Additionally, the contribution of the diffuse radio lobes to the low-resolution flux densities cannot be measured with the current observations.
Furthermore, the exact flux ratio between the two hotspots cannot be determined unambiguously because of the limited resolution and sensitivity throughout all observations.
As a consequence, the true spectral shapes of NW and SE hotspots will change together with the flux densities of the lobes and flux densities ratios at $\nu\leq8.2$ GHz.
Hence, the obtained synchrotron age is just an order of magnitude estimation.

We show the distributions of the best-fit parameters under the assumptions that, at low frequencies, the fractional contamination of diffuse emissions to the observed flux densities ranges from 1 to 55 per cent by a step of 1, and flux ratios vary from 4 to 15 by a step of 1, in Fig.~\ref{fig:synchrofit statistics}.
Based on the current estimate of $\tnr{B}{eq}$, unless $\log\tnr{\nu}{br}$ is smaller than 8.1 (corresponding to $\sim150$ MHz), the age cannot exceed $10^5$ yr.
A very low break frequency smaller than 150 MHz is not expected for all the fittings, suggesting that the conclusion of a young radio source is not sensitive to the limited number of data points and uncertain contaminations and flux ratios.
Although a non-zero remnant fraction is favored in most fittings, when the SE hotspot is fitted only up to 44 GHz, a zero remnant fraction does happen in some cases.
These are unavoidable uncertainties in this work due to the limited resolved low-frequency observations and possibly contaminated 237 GHz data, by which we cannot constrain the real spectral shapes.
Future observations are essentially required to decompose the ALMA 237 GHz data and fill the gap between 44 and 237 GHz to confirm whether the curvature is robust and in need of a non-zero remnant fraction to interpret.
We, therefore, emphasize the importance of high-frequency observations which can help us investigate the aging problem of radio sources.
We also note the importance of multi-frequency data since the limited number of data points can dilute the curvature of the radio SED, leading to further uncertainties in the estimation of the remnant fraction.

\section{Conclusion}
In this work, we have studied the synchrotron radiation from the radio hotspots in the Dragonfly galaxy, a hyper-luminous infrared galaxy at $z=1.92$ using joint VLA and ALMA observations.
Our major findings are as follows:
\begin{enumerate}[leftmargin=*]

\item The NW hotspot, SE hotspot, and radio core constitute the synchrotron radiation from the Dragonfly galaxy and the SE hotspot dominates the observed flux density.
The synchrotron source has a projected linear size of $\sim13$ kpc and $\alpha<-0.5$ at $\nu_\mathrm{obs}\geq365$ MHz, being classified as a CSS source.

\item ALMA Band 6 observation catches a sub-component offset from the SE galaxy by $\sim1$ kpc and coincided perfectly with the location of the SE hotspot.
This sub-component may be a mixture of different emission mechanisms, including synchrotron radiation, thermal bremsstrahlung, and Rayleigh-Jeans tail.

\item The NW hotspot has a synchrotron age $t_\mathrm{s}\sim2-3\times10^5$ yr, while the SE one can vary amongst $t_\mathrm{s}\sim2-4\times10^5$ yr.
These age estimates agree with typical orders of magnitude observed in CSS sources related to radio AGNs that are robustly young.
Furthermore, the fittings of both hotspots may indicate that the radio jets have been switched off at the observation timing, or that the RLAGN has past radio activities and re-launched radio jets that are not powerful enough to alter the energy density of high-frequency synchrotron-emitting electrons.
This suggests that this RLAGN may have transient or intermittent activities because the central SMBH of the AGN residing in the NW galaxy is possibly in a fast transition phase.

\item The NW hotspot has an $\alpha_\mathrm{inj}=-0.59$ and the SE one has $\alpha_\mathrm{inj}=-0.51$ in fittings without ALMA 237 GHz data, both consistent with the values expected for the first-order Fermi acceleration.
When the 237 GHz data is considered, the SE hotspot has $\alpha_\mathrm{inj}=-0.82$.
If this value is correct, it may indicate that the particle acceleration in the SE hotspot may not simply be explained by the first-order Fermi acceleration for a non-relativistic strong shock in a uniform-density ISM, reflecting an environmental difference between the NW and SE hotspots.
However, because of the limited resolution, the possibility that this difference is a result of the contaminated 237 GHz and low-frequency data exists.

\item The observed flux ratios between the SE and NW spots indicate that the intrinsic flux density of the SE hotspot has been Doppler boosted as a result of the smaller inclination angle of the radio jet relative to the LOS, while the NW one dimmed.
The SE (NW) hotspot then corresponds to the approaching (receding) jet.
The advance speed of the jet head ranges from $\sim$0.1c -- 0.3c, in agreement with the mildly relativistic jet case observed in typical FRII galaxies.

\item If the sub-component identified in ALMA Cycle 6 observation indicates an in situ jet-ISM interaction, the jet can drive a massive molecular gas outflow within its lifetime and excite the CO gas to higher rotational transition levels, providing an explanation for the extended and diffuse CO(6-5) line emission on the west side of the SE galaxy, as well as the steep $\alpha_\mathrm{inj}$ found in the SE hotspot fitted by the CI model.
\end{enumerate}

The remnant fraction is robust for a reproduction of the curvature of the observed radio SED, suggesting that this RLAGN may have transient or intermittent activities.
Considering also its young age, this suggests that high-redshift RLAGNs may have short duty cycles because of the stochastic accretion flows onto SMBHs.
Still, this requires a larger sample to confirm.
We note the importance of high-frequency observations to investigate the behavior of radio-loud AGNs at high redshifts to understand when they interact with their ambient medium, such that we can gain a more complete view of AGN-host galaxy co-evolution.

The sub-component adjacent to the SE galaxy may be a chance alignment.
Otherwise, this HzRG is a merging galaxy in which the radio jet launched from the AGN interacts with the merging pair of the AGN host galaxy. 
There are only a few cases of such kind of a system observed at lower redshifts \citep[e.g.,][]{2022MNRAS.517L..86H} and this galaxy may be the first one identified at high redshifts.
Therefore, subsequent observations, including both radio and optical observations, are required to confirm this scenario, as well as to decompose the different emission mechanisms at 237 GHz.

\section*{Acknowledgements}
We thank the staff in ALMA and NRAO helpdesk for their kind help in data calibration and reduction.
We thank Bjorn Emonts and Sophie Lebowitz for sharing their paper before it becomes public.
We also thank Niinuma Kotaro for his comments.
This paper makes use of the following VLA data: VLA/15A-316 and VLA/17B-444.
This paper makes use of the following ALMA data: ADS/JAO.ALMA\#2016.1.01417.S and ADS/JAO.ALMA\#2018.1.00293.S.
Data analysis was carried out on the Multi-wavelength Data Analysis System operated by the Astronomy Data Center (ADC), National Astronomical Observatory of Japan.
AKI, YS, and YF are supported by NAOJ ALMA Scientific Research Grant Numbers 2020-16B.
ALMA is a partnership of ESO (representing its member states), NSF (USA) and NINS (Japan), together with NRC (Canada), MOST and ASIAA (Taiwan), and KASI (Republic of Korea), in cooperation with the Republic of Chile. 
The Joint ALMA Observatory is operated by ESO, AUI/NRAO and NAOJ. 
The National Radio Astronomy Observatory is a facility of the National Science Foundation operated under cooperative agreement by Associated Universities, Inc.

\textit{Software:} 
\textsc{python} \citep{10.5555/1593511}, 
\textsc{matplotlib} \citep{Hunter:2007}, 
\textsc{astropy} \citep{astropy:2013,astropy:2018,astropy:2022},
\textsc{synchrofit} \citep{2018MNRAS.476.2522T,2018MNRAS.474.3361T}, 
\textsc{carta} \citep[Cube Analysis and Rendering Tool for Astronomy,][]{angus_comrie_2021_4905459},
\textsc{numpy} \citep{harris2020array},
and \textsc{scipy} \citep{2020SciPy-NMeth}.

\section*{Data Availability}
The ALMA data used in this work are publicly available at https://almascience.nao.ac.jp/aq/.
The VLA data used in this work are publicly available at https://data.nrao.edu.

\bibliographystyle{mnras}
\bibliography{dragonfly_jet}

\begin{thebibliography}{}
\makeatletter
\relax
\def\mn@urlcharsother{\let\do\@makeother \do\$\do\&\do\#\do\^\do\_\do\%\do\~}
\def\mn@doi{\begingroup\mn@urlcharsother \@ifnextchar [ {\mn@doi@}
  {\mn@doi@[]}}
\def\mn@doi@[#1]#2{\def\@tempa{#1}\ifx\@tempa\@empty \href
  {http://dx.doi.org/#2} {doi:#2}\else \href {http://dx.doi.org/#2} {#1}\fi
  \endgroup}
\def\mn@eprint#1#2{\mn@eprint@#1:#2::\@nil}
\def\mn@eprint@arXiv#1{\href {http://arxiv.org/abs/#1} {{\tt arXiv:#1}}}
\def\mn@eprint@dblp#1{\href {http://dblp.uni-trier.de/rec/bibtex/#1.xml}
  {dblp:#1}}
\def\mn@eprint@#1:#2:#3:#4\@nil{\def\@tempa {#1}\def\@tempb {#2}\def\@tempc
  {#3}\ifx \@tempc \@empty \let \@tempc \@tempb \let \@tempb \@tempa \fi \ifx
  \@tempb \@empty \def\@tempb {arXiv}\fi \@ifundefined
  {mn@eprint@\@tempb}{\@tempb:\@tempc}{\expandafter \expandafter \csname
  mn@eprint@\@tempb\endcsname \expandafter{\@tempc}}}

\bibitem[\protect\citeauthoryear{{An} et~al.,}{{An}
  et~al.}{2012}]{2012ApJS..198....5A}
{An} T.,  et~al., 2012, \mn@doi [\apjs] {10.1088/0067-0049/198/1/5}, \href
  {https://ui.adsabs.harvard.edu/abs/2012ApJS..198....5A} {198, 5}

\bibitem[\protect\citeauthoryear{{Araudo}, {Bell}, {Matthews}  \&
  {Blundell}}{{Araudo} et~al.}{2019}]{2019EPJWC.21004006A}
{Araudo} A.~T.,  {Bell} A.~R.,  {Matthews} J.,   {Blundell} K.,  2019, in
  European Physical Journal Web of Conferences. p. 04006,
  \mn@doi{10.1051/epjconf/201921004006}

\bibitem[\protect\citeauthoryear{{Astropy Collaboration} et~al.,}{{Astropy
  Collaboration} et~al.}{2013}]{astropy:2013}
{Astropy Collaboration} et~al., 2013, \mn@doi [\aap]
  {10.1051/0004-6361/201322068}, \href
  {http://adsabs.harvard.edu/abs/2013A%26A...558A..33A} {558, A33}

\bibitem[\protect\citeauthoryear{{Astropy Collaboration} et~al.,}{{Astropy
  Collaboration} et~al.}{2018}]{astropy:2018}
{Astropy Collaboration} et~al., 2018, \mn@doi [\aj] {10.3847/1538-3881/aabc4f},
  \href {https://ui.adsabs.harvard.edu/abs/2018AJ....156..123A} {156, 123}

\bibitem[\protect\citeauthoryear{{Astropy Collaboration} et~al.,}{{Astropy
  Collaboration} et~al.}{2022}]{astropy:2022}
{Astropy Collaboration} et~al., 2022, \mn@doi [apj] {10.3847/1538-4357/ac7c74},
  \href {https://ui.adsabs.harvard.edu/abs/2022ApJ...935..167A} {935, 167}

\bibitem[\protect\citeauthoryear{{Bambic}, {Russell}, {Reynolds}, {Fabian},
  {McNamara}  \& {Nulsen}}{{Bambic} et~al.}{2023}]{2023arXiv230111937B}
{Bambic} C.~J.,  {Russell} H.~R.,  {Reynolds} C.~S.,  {Fabian} A.~C.,
  {McNamara} B.~R.,   {Nulsen} P.~E.~J.,  2023, \mn@doi [arXiv e-prints]
  {10.48550/arXiv.2301.11937}, \href
  {https://ui.adsabs.harvard.edu/abs/2023arXiv230111937B} {p. arXiv:2301.11937}

\bibitem[\protect\citeauthoryear{{Best}, {Kauffmann}, {Heckman}, {Brinchmann},
  {Charlot}, {Ivezi{\'c}}  \& {White}}{{Best}
  et~al.}{2005}]{2005MNRAS.362...25B}
{Best} P.~N.,  {Kauffmann} G.,  {Heckman} T.~M.,  {Brinchmann} J.,  {Charlot}
  S.,  {Ivezi{\'c}} {\v{Z}}.,   {White} S.~D.~M.,  2005, \mn@doi [\mnras]
  {10.1111/j.1365-2966.2005.09192.x}, \href
  {https://ui.adsabs.harvard.edu/abs/2005MNRAS.362...25B} {362, 25}

\bibitem[\protect\citeauthoryear{{Bondi}, {Ciliegi}, {Schinnerer},
  {Smol{\v{c}}i{\'c}}, {Jahnke}, {Carilli}  \& {Zamorani}}{{Bondi}
  et~al.}{2008}]{2008ApJ...681.1129B}
{Bondi} M.,  {Ciliegi} P.,  {Schinnerer} E.,  {Smol{\v{c}}i{\'c}} V.,  {Jahnke}
  K.,  {Carilli} C.,   {Zamorani} G.,  2008, \mn@doi [\apj] {10.1086/589324},
  \href {https://ui.adsabs.harvard.edu/abs/2008ApJ...681.1129B} {681, 1129}

\bibitem[\protect\citeauthoryear{{Callingham} et~al.,}{{Callingham}
  et~al.}{2017}]{2017ApJ...836..174C}
{Callingham} J.~R.,  et~al., 2017, \mn@doi [\apj]
  {10.3847/1538-4357/836/2/174}, \href
  {https://ui.adsabs.harvard.edu/abs/2017ApJ...836..174C} {836, 174}

\bibitem[\protect\citeauthoryear{{Carilli}, {Perley}, {Dreher}  \&
  {Leahy}}{{Carilli} et~al.}{1991}]{1991ApJ...383..554C}
{Carilli} C.~L.,  {Perley} R.~A.,  {Dreher} J.~W.,   {Leahy} J.~P.,  1991,
  \mn@doi [\apj] {10.1086/170813}, \href
  {https://ui.adsabs.harvard.edu/abs/1991ApJ...383..554C} {383, 554}

\bibitem[\protect\citeauthoryear{{Cavagnolo}, {McNamara}, {Nulsen}, {Carilli},
  {Jones}  \& {B{\^\i}rzan}}{{Cavagnolo} et~al.}{2010}]{2010ApJ...720.1066C}
{Cavagnolo} K.~W.,  {McNamara} B.~R.,  {Nulsen} P.~E.~J.,  {Carilli} C.~L.,
  {Jones} C.,   {B{\^\i}rzan} L.,  2010, \mn@doi [\apj]
  {10.1088/0004-637X/720/2/1066}, \href
  {https://ui.adsabs.harvard.edu/abs/2010ApJ...720.1066C} {720, 1066}

\bibitem[\protect\citeauthoryear{Comrie et~al.,}{Comrie
  et~al.}{2021}]{angus_comrie_2021_4905459}
Comrie A.,  et~al., 2021, {CARTA: The Cube Analysis and Rendering Tool for
  Astronomy}, \mn@doi{10.5281/zenodo.4905459}, \url
  {https://doi.org/10.5281/zenodo.4905459}

\bibitem[\protect\citeauthoryear{{Condon}}{{Condon}}{1992}]{1992ARA&A..30..575C}
{Condon} J.~J.,  1992, \mn@doi [\araa] {10.1146/annurev.aa.30.090192.003043},
  \href {https://ui.adsabs.harvard.edu/abs/1992ARA&A..30..575C} {30, 575}

\bibitem[\protect\citeauthoryear{{Condon} \& {Ransom}}{{Condon} \&
  {Ransom}}{2016}]{2016era..book.....C}
{Condon} J.~J.,  {Ransom} S.~M.,  2016, {Essential Radio Astronomy}

\bibitem[\protect\citeauthoryear{{De Breuck}, {van Breugel}, {R{\"o}ttgering}
  \& {Miley}}{{De Breuck} et~al.}{2000}]{2000A&AS..143..303D}
{De Breuck} C.,  {van Breugel} W.,  {R{\"o}ttgering} H.~J.~A.,   {Miley} G.,
  2000, \mn@doi [\aaps] {10.1051/aas:2000181}, \href
  {https://ui.adsabs.harvard.edu/abs/2000A&AS..143..303D} {143, 303}

\bibitem[\protect\citeauthoryear{{Dermer} \& {Menon}}{{Dermer} \&
  {Menon}}{2009}]{2009herb.book.....D}
{Dermer} C.~D.,  {Menon} G.,  2009, {High Energy Radiation from Black Holes:
  Gamma Rays, Cosmic Rays, and Neutrinos}

\bibitem[\protect\citeauthoryear{{Dicken} et~al.,}{{Dicken}
  et~al.}{2012}]{2012ApJ...745..172D}
{Dicken} D.,  et~al., 2012, \mn@doi [\apj] {10.1088/0004-637X/745/2/172}, \href
  {https://ui.adsabs.harvard.edu/abs/2012ApJ...745..172D} {745, 172}

\bibitem[\protect\citeauthoryear{{Drouart} et~al.,}{{Drouart}
  et~al.}{2014}]{2014A&A...566A..53D}
{Drouart} G.,  et~al., 2014, \mn@doi [\aap] {10.1051/0004-6361/201323310},
  \href {https://ui.adsabs.harvard.edu/abs/2014A&A...566A..53D} {566, A53}

\bibitem[\protect\citeauthoryear{{Duffy} \& {Blundell}}{{Duffy} \&
  {Blundell}}{2012}]{2012MNRAS.421..108D}
{Duffy} P.,  {Blundell} K.~M.,  2012, \mn@doi [\mnras]
  {10.1111/j.1365-2966.2011.20239.x}, \href
  {https://ui.adsabs.harvard.edu/abs/2012MNRAS.421..108D} {421, 108}

\bibitem[\protect\citeauthoryear{{Emonts} et~al.,}{{Emonts}
  et~al.}{2011}]{2011ApJ...734L..25E}
{Emonts} B.~H.~C.,  et~al., 2011, \mn@doi [\apjl]
  {10.1088/2041-8205/734/1/L25}, \href
  {https://ui.adsabs.harvard.edu/abs/2011ApJ...734L..25E} {734, L25}

\bibitem[\protect\citeauthoryear{{Emonts} et~al.,}{{Emonts}
  et~al.}{2015a}]{2015MNRAS.451.1025E}
{Emonts} B.~H.~C.,  et~al., 2015a, \mn@doi [\mnras] {10.1093/mnras/stv930},
  \href {https://ui.adsabs.harvard.edu/abs/2015MNRAS.451.1025E} {451, 1025}

\bibitem[\protect\citeauthoryear{{Emonts} et~al.,}{{Emonts}
  et~al.}{2015b}]{2015A&A...584A..99E}
{Emonts} B.~H.~C.,  et~al., 2015b, \mn@doi [\aap]
  {10.1051/0004-6361/201526090}, \href
  {https://ui.adsabs.harvard.edu/abs/2015A&A...584A..99E} {584, A99}

\bibitem[\protect\citeauthoryear{{Emonts} et~al.,}{{Emonts}
  et~al.}{2023}]{2023arXiv230317484E}
{Emonts} B. H.~C.,  et~al., 2023, \mn@doi [arXiv e-prints]
  {10.48550/arXiv.2303.17484}, \href
  {https://ui.adsabs.harvard.edu/abs/2023arXiv230317484E} {p. arXiv:2303.17484}

\bibitem[\protect\citeauthoryear{{Garc{\'\i}a-Burillo}
  et~al.,}{{Garc{\'\i}a-Burillo} et~al.}{2019}]{2019A&A...632A..61G}
{Garc{\'\i}a-Burillo} S.,  et~al., 2019, \mn@doi [\aap]
  {10.1051/0004-6361/201936606}, \href
  {https://ui.adsabs.harvard.edu/abs/2019A&A...632A..61G} {632, A61}

\bibitem[\protect\citeauthoryear{{G{\"u}rkan} et~al.,}{{G{\"u}rkan}
  et~al.}{2015}]{2015MNRAS.452.3776G}
{G{\"u}rkan} G.,  et~al., 2015, \mn@doi [\mnras] {10.1093/mnras/stv1502}, \href
  {https://ui.adsabs.harvard.edu/abs/2015MNRAS.452.3776G} {452, 3776}

\bibitem[\protect\citeauthoryear{{Hardcastle} \& {Croston}}{{Hardcastle} \&
  {Croston}}{2020}]{2020NewAR..8801539H}
{Hardcastle} M.~J.,  {Croston} J.~H.,  2020, \mn@doi [\nar]
  {10.1016/j.newar.2020.101539}, \href
  {https://ui.adsabs.harvard.edu/abs/2020NewAR..8801539H} {88, 101539}

\bibitem[\protect\citeauthoryear{Harris et~al.,}{Harris
  et~al.}{2020}]{harris2020array}
Harris C.~R.,  et~al., 2020, \mn@doi [Nature] {10.1038/s41586-020-2649-2}, 585,
  357

\bibitem[\protect\citeauthoryear{{Heckman} \& {Best}}{{Heckman} \&
  {Best}}{2014}]{2014ARA&A..52..589H}
{Heckman} T.~M.,  {Best} P.~N.,  2014, \mn@doi [\araa]
  {10.1146/annurev-astro-081913-035722}, \href
  {https://ui.adsabs.harvard.edu/abs/2014ARA&A..52..589H} {52, 589}

\bibitem[\protect\citeauthoryear{{Hota} et~al.,}{{Hota}
  et~al.}{2022}]{2022MNRAS.517L..86H}
{Hota} A.,  et~al., 2022, \mn@doi [\mnras] {10.1093/mnrasl/slac116}, \href
  {https://ui.adsabs.harvard.edu/abs/2022MNRAS.517L..86H} {517, L86}

\bibitem[\protect\citeauthoryear{Hunter}{Hunter}{2007}]{Hunter:2007}
Hunter J.~D.,  2007, \mn@doi [Computing in Science \& Engineering]
  {10.1109/MCSE.2007.55}, 9, 90

\bibitem[\protect\citeauthoryear{{Jaffe} \& {Perola}}{{Jaffe} \&
  {Perola}}{1973}]{1973A&A....26..423J}
{Jaffe} W.~J.,  {Perola} G.~C.,  1973, \aap, \href
  {https://ui.adsabs.harvard.edu/abs/1973A&A....26..423J} {26, 423}

\bibitem[\protect\citeauthoryear{{Jeyakumar} \& {Saikia}}{{Jeyakumar} \&
  {Saikia}}{2000}]{2000MNRAS.311..397J}
{Jeyakumar} S.,  {Saikia} D.~J.,  2000, \mn@doi [\mnras]
  {10.1046/j.1365-8711.2000.03063.x}, \href
  {https://ui.adsabs.harvard.edu/abs/2000MNRAS.311..397J} {311, 397}

\bibitem[\protect\citeauthoryear{{Kappes}, {Perucho}, {Kadler}, {Burd},
  {Vega-Garc{\'\i}a}  \& {Br{\"u}ggen}}{{Kappes}
  et~al.}{2019}]{2019A&A...631A..49K}
{Kappes} A.,  {Perucho} M.,  {Kadler} M.,  {Burd} P.~R.,  {Vega-Garc{\'\i}a}
  L.,   {Br{\"u}ggen} M.,  2019, \mn@doi [\aap] {10.1051/0004-6361/201936164},
  \href {https://ui.adsabs.harvard.edu/abs/2019A&A...631A..49K} {631, A49}

\bibitem[\protect\citeauthoryear{{Kardashev}}{{Kardashev}}{1962}]{1962SvA.....6..317K}
{Kardashev} N.~S.,  1962, \sovast, \href
  {https://ui.adsabs.harvard.edu/abs/1962SvA.....6..317K} {6, 317}

\bibitem[\protect\citeauthoryear{{Kellermann}, {Sramek}, {Schmidt}, {Shaffer}
  \& {Green}}{{Kellermann} et~al.}{1989}]{1989AJ.....98.1195K}
{Kellermann} K.~I.,  {Sramek} R.,  {Schmidt} M.,  {Shaffer} D.~B.,   {Green}
  R.,  1989, \mn@doi [\aj] {10.1086/115207}, \href
  {https://ui.adsabs.harvard.edu/abs/1989AJ.....98.1195K} {98, 1195}

\bibitem[\protect\citeauthoryear{{Kellermann}, {Sramek}, {Schmidt}, {Green}  \&
  {Shaffer}}{{Kellermann} et~al.}{1994}]{1994AJ....108.1163K}
{Kellermann} K.~I.,  {Sramek} R.~A.,  {Schmidt} M.,  {Green} R.~F.,   {Shaffer}
  D.~B.,  1994, \mn@doi [\aj] {10.1086/117145}, \href
  {https://ui.adsabs.harvard.edu/abs/1994AJ....108.1163K} {108, 1163}

\bibitem[\protect\citeauthoryear{{Kellermann}, {Condon}, {Kimball}, {Perley}
  \& {Ivezi{\'c}}}{{Kellermann} et~al.}{2016}]{2016ApJ...831..168K}
{Kellermann} K.~I.,  {Condon} J.~J.,  {Kimball} A.~E.,  {Perley} R.~A.,
  {Ivezi{\'c}} {\v{Z}}.,  2016, \mn@doi [\apj] {10.3847/0004-637X/831/2/168},
  \href {https://ui.adsabs.harvard.edu/abs/2016ApJ...831..168K} {831, 168}

\bibitem[\protect\citeauthoryear{{Maccagni} et~al.,}{{Maccagni}
  et~al.}{2020}]{2020A&A...634A...9M}
{Maccagni} F.~M.,  et~al., 2020, \mn@doi [\aap] {10.1051/0004-6361/201936867},
  \href {https://ui.adsabs.harvard.edu/abs/2020A&A...634A...9M} {634, A9}

\bibitem[\protect\citeauthoryear{{McMullin}, {Waters}, {Schiebel}, {Young}  \&
  {Golap}}{{McMullin} et~al.}{2007}]{2007ASPC..376..127M}
{McMullin} J.~P.,  {Waters} B.,  {Schiebel} D.,  {Young} W.,   {Golap} K.,
  2007, in {Shaw} R.~A.,  {Hill} F.,   {Bell} D.~J.,  eds,  Astronomical
  Society of the Pacific Conference Series Vol. 376, Astronomical Data Analysis
  Software and Systems XVI. p.~127

\bibitem[\protect\citeauthoryear{{Miley}}{{Miley}}{1980}]{1980ARA&A..18..165M}
{Miley} G.,  1980, \mn@doi [\araa] {10.1146/annurev.aa.18.090180.001121}, \href
  {https://ui.adsabs.harvard.edu/abs/1980ARA&A..18..165M} {18, 165}

\bibitem[\protect\citeauthoryear{{Miley} \& {De Breuck}}{{Miley} \& {De
  Breuck}}{2008}]{2008A&ARv..15...67M}
{Miley} G.,  {De Breuck} C.,  2008, \mn@doi [\aapr]
  {10.1007/s00159-007-0008-z}, \href
  {https://ui.adsabs.harvard.edu/abs/2008A&ARv..15...67M} {15, 67}

\bibitem[\protect\citeauthoryear{{Mukherjee}, {Bicknell}, {Sutherland}  \&
  {Wagner}}{{Mukherjee} et~al.}{2016}]{2016MNRAS.461..967M}
{Mukherjee} D.,  {Bicknell} G.~V.,  {Sutherland} R.,   {Wagner} A.,  2016,
  \mn@doi [\mnras] {10.1093/mnras/stw1368}, \href
  {https://ui.adsabs.harvard.edu/abs/2016MNRAS.461..967M} {461, 967}

\bibitem[\protect\citeauthoryear{{Mukherjee}, {Bicknell}, {Wagner},
  {Sutherland}  \& {Silk}}{{Mukherjee} et~al.}{2018}]{2018MNRAS.479.5544M}
{Mukherjee} D.,  {Bicknell} G.~V.,  {Wagner} A.~Y.,  {Sutherland} R.~S.,
  {Silk} J.,  2018, \mn@doi [\mnras] {10.1093/mnras/sty1776}, \href
  {https://ui.adsabs.harvard.edu/abs/2018MNRAS.479.5544M} {479, 5544}

\bibitem[\protect\citeauthoryear{{Murgia}}{{Murgia}}{2003}]{2003PASA...20...19M}
{Murgia} M.,  2003, \mn@doi [\pasa] {10.1071/AS02033}, \href
  {https://ui.adsabs.harvard.edu/abs/2003PASA...20...19M} {20, 19}

\bibitem[\protect\citeauthoryear{{Murgia} et~al.,}{{Murgia}
  et~al.}{2011}]{2011A&A...526A.148M}
{Murgia} M.,  et~al., 2011, \mn@doi [\aap] {10.1051/0004-6361/201015302}, \href
  {https://ui.adsabs.harvard.edu/abs/2011A&A...526A.148M} {526, A148}

\bibitem[\protect\citeauthoryear{{Nesvadba}, {Lehnert}, {Eisenhauer},
  {Gilbert}, {Tecza}  \& {Abuter}}{{Nesvadba}
  et~al.}{2006}]{2006ApJ...650..693N}
{Nesvadba} N.~P.~H.,  {Lehnert} M.~D.,  {Eisenhauer} F.,  {Gilbert} A.,
  {Tecza} M.,   {Abuter} R.,  2006, \mn@doi [\apj] {10.1086/507266}, \href
  {https://ui.adsabs.harvard.edu/abs/2006ApJ...650..693N} {650, 693}

\bibitem[\protect\citeauthoryear{{Nesvadba}, {Lehnert}, {De Breuck}, {Gilbert}
  \& {van Breugel}}{{Nesvadba} et~al.}{2007}]{2007A&A...475..145N}
{Nesvadba} N.~P.~H.,  {Lehnert} M.~D.,  {De Breuck} C.,  {Gilbert} A.,   {van
  Breugel} W.,  2007, \mn@doi [\aap] {10.1051/0004-6361:20078175}, \href
  {https://ui.adsabs.harvard.edu/abs/2007A&A...475..145N} {475, 145}

\bibitem[\protect\citeauthoryear{{Nesvadba}, {Drouart}, {De Breuck}, {Best},
  {Seymour}  \& {Vernet}}{{Nesvadba} et~al.}{2017}]{2017A&A...600A.121N}
{Nesvadba} N.~P.~H.,  {Drouart} G.,  {De Breuck} C.,  {Best} P.,  {Seymour} N.,
    {Vernet} J.,  2017, \mn@doi [\aap] {10.1051/0004-6361/201629357}, \href
  {https://ui.adsabs.harvard.edu/abs/2017A&A...600A.121N} {600, A121}

\bibitem[\protect\citeauthoryear{{O'Dea}}{{O'Dea}}{1998}]{1998PASP..110..493O}
{O'Dea} C.~P.,  1998, \mn@doi [\pasp] {10.1086/316162}, \href
  {https://ui.adsabs.harvard.edu/abs/1998PASP..110..493O} {110, 493}

\bibitem[\protect\citeauthoryear{{O'Dea} \& {Baum}}{{O'Dea} \&
  {Baum}}{1997}]{1997AJ....113..148O}
{O'Dea} C.~P.,  {Baum} S.~A.,  1997, \mn@doi [\aj] {10.1086/118241}, \href
  {https://ui.adsabs.harvard.edu/abs/1997AJ....113..148O} {113, 148}

\bibitem[\protect\citeauthoryear{{O'Dea} \& {Saikia}}{{O'Dea} \&
  {Saikia}}{2021}]{2021A&ARv..29....3O}
{O'Dea} C.~P.,  {Saikia} D.~J.,  2021, \mn@doi [\aapr]
  {10.1007/s00159-021-00131-w}, \href
  {https://ui.adsabs.harvard.edu/abs/2021A&ARv..29....3O} {29, 3}

\bibitem[\protect\citeauthoryear{{O'Dea}, {Daly}, {Kharb}, {Freeman}  \&
  {Baum}}{{O'Dea} et~al.}{2009}]{2009A&A...494..471O}
{O'Dea} C.~P.,  {Daly} R.~A.,  {Kharb} P.,  {Freeman} K.~A.,   {Baum} S.~A.,
  2009, \mn@doi [\aap] {10.1051/0004-6361:200809416}, \href
  {https://ui.adsabs.harvard.edu/abs/2009A&A...494..471O} {494, 471}

\bibitem[\protect\citeauthoryear{{Pacholczyk}}{{Pacholczyk}}{1970}]{1970ranp.book.....P}
{Pacholczyk} A.~G.,  1970, {Radio astrophysics. Nonthermal processes in
  galactic and extragalactic sources}

\bibitem[\protect\citeauthoryear{{Patil} et~al.,}{{Patil}
  et~al.}{2020}]{2020ApJ...896...18P}
{Patil} P.,  et~al., 2020, \mn@doi [\apj] {10.3847/1538-4357/ab9011}, \href
  {https://ui.adsabs.harvard.edu/abs/2020ApJ...896...18P} {896, 18}

\bibitem[\protect\citeauthoryear{{Pentericci}, {Van Reeven}, {Carilli},
  {R{\"o}ttgering}  \& {Miley}}{{Pentericci} et~al.}{2000}]{2000AAS..145..121P}
{Pentericci} L.,  {Van Reeven} W.,  {Carilli} C.~L.,  {R{\"o}ttgering}
  H.~J.~A.,   {Miley} G.~K.,  2000, \mn@doi [\aaps] {10.1051/aas:2000104},
  \href {https://ui.adsabs.harvard.edu/abs/2000A&AS..145..121P} {145, 121}

\bibitem[\protect\citeauthoryear{{Planck Collaboration} et~al.,}{{Planck
  Collaboration} et~al.}{2016}]{2016A&A...594A..13P}
{Planck Collaboration} et~al., 2016, \mn@doi [\aap]
  {10.1051/0004-6361/201525830}, \href
  {https://ui.adsabs.harvard.edu/abs/2016A&A...594A..13P} {594, A13}

\bibitem[\protect\citeauthoryear{{Prandoni}, {Gregorini}, {Parma}, {de Ruiter},
  {Vettolani}, {Wieringa}  \& {Ekers}}{{Prandoni}
  et~al.}{2001}]{2001A&A...365..392P}
{Prandoni} I.,  {Gregorini} L.,  {Parma} P.,  {de Ruiter} H.~R.,  {Vettolani}
  G.,  {Wieringa} M.~H.,   {Ekers} R.~D.,  2001, \mn@doi [\aap]
  {10.1051/0004-6361:20000142}, \href
  {https://ui.adsabs.harvard.edu/abs/2001A&A...365..392P} {365, 392}

\bibitem[\protect\citeauthoryear{{Saxena}, {R{\"o}ttgering}  \&
  {Rigby}}{{Saxena} et~al.}{2017}]{2017MNRAS.469.4083S}
{Saxena} A.,  {R{\"o}ttgering} H.~J.~A.,   {Rigby} E.~E.,  2017, \mn@doi
  [\mnras] {10.1093/mnras/stx1150}, \href
  {https://ui.adsabs.harvard.edu/abs/2017MNRAS.469.4083S} {469, 4083}

\bibitem[\protect\citeauthoryear{{Saxena} et~al.,}{{Saxena}
  et~al.}{2018}]{2018MNRAS.480.2733S}
{Saxena} A.,  et~al., 2018, \mn@doi [\mnras] {10.1093/mnras/sty1996}, \href
  {https://ui.adsabs.harvard.edu/abs/2018MNRAS.480.2733S} {480, 2733}

\bibitem[\protect\citeauthoryear{Snell, Kurtz  \& Marr}{Snell
  et~al.}{2019}]{snell2019fundamentals}
Snell R.,  Kurtz S.,   Marr J.,  2019, Fundamentals of Radio Astronomy:
  Astrophysics.
Series in Astronomy and Astrophysics, CRC Press, \url
  {https://books.google.co.jp/books?id=7XyUDwAAQBAJ}

\bibitem[\protect\citeauthoryear{{Sotnikova}, {Mufakharov}, {Majorova},
  {Mingaliev}, {Udovitskii}, {Bursov}  \& {Semenova}}{{Sotnikova}
  et~al.}{2019}]{2019AstBu..74..348S}
{Sotnikova} Y.~V.,  {Mufakharov} T.~V.,  {Majorova} E.~K.,  {Mingaliev} M.~G.,
  {Udovitskii} R.~Y.,  {Bursov} N.~N.,   {Semenova} T.~A.,  2019, \mn@doi
  [Astrophysical Bulletin] {10.1134/S1990341319040023}, \href
  {https://ui.adsabs.harvard.edu/abs/2019AstBu..74..348S} {74, 348}

\bibitem[\protect\citeauthoryear{{THE CASA TEAM} et~al.,}{{THE CASA TEAM}
  et~al.}{2022}]{2022arXiv221002276T}
{THE CASA TEAM} et~al., 2022, arXiv e-prints, \href
  {https://ui.adsabs.harvard.edu/abs/2022arXiv221002276T} {p. arXiv:2210.02276}

\bibitem[\protect\citeauthoryear{{Talbot}, {Sijacki}  \& {Bourne}}{{Talbot}
  et~al.}{2022}]{2022MNRAS.514.4535T}
{Talbot} R.~Y.,  {Sijacki} D.,   {Bourne} M.~A.,  2022, \mn@doi [\mnras]
  {10.1093/mnras/stac1566}, \href
  {https://ui.adsabs.harvard.edu/abs/2022MNRAS.514.4535T} {514, 4535}

\bibitem[\protect\citeauthoryear{{Turner}}{{Turner}}{2018}]{2018MNRAS.476.2522T}
{Turner} R.~J.,  2018, \mn@doi [\mnras] {10.1093/mnras/sty433}, \href
  {https://ui.adsabs.harvard.edu/abs/2018MNRAS.476.2522T} {476, 2522}

\bibitem[\protect\citeauthoryear{{Turner}, {Shabala}  \& {Krause}}{{Turner}
  et~al.}{2018}]{2018MNRAS.474.3361T}
{Turner} R.~J.,  {Shabala} S.~S.,   {Krause} M. G.~H.,  2018, \mn@doi [\mnras]
  {10.1093/mnras/stx2947}, \href
  {https://ui.adsabs.harvard.edu/abs/2018MNRAS.474.3361T} {474, 3361}

\bibitem[\protect\citeauthoryear{Van~Rossum \& Drake}{Van~Rossum \&
  Drake}{2009}]{10.5555/1593511}
Van~Rossum G.,  Drake F.~L.,  2009, Python 3 Reference Manual.
CreateSpace, Scotts Valley, CA

\bibitem[\protect\citeauthoryear{Virtanen et~al.,}{Virtanen
  et~al.}{2020}]{2020SciPy-NMeth}
Virtanen P.,  et~al., 2020, \mn@doi [Nature Methods]
  {10.1038/s41592-019-0686-2}, \href {https://rdcu.be/b08Wh} {17, 261}

\bibitem[\protect\citeauthoryear{{Vollmer} et~al.,}{{Vollmer}
  et~al.}{2010}]{2010AA...511A..53V}
{Vollmer} B.,  et~al., 2010, \mn@doi [\aap] {10.1051/0004-6361/200913460},
  \href {https://ui.adsabs.harvard.edu/abs/2010A&A...511A..53V} {511, A53}

\bibitem[\protect\citeauthoryear{{Wagner}, {Bicknell}  \& {Umemura}}{{Wagner}
  et~al.}{2012}]{2012ApJ...757..136W}
{Wagner} A.~Y.,  {Bicknell} G.~V.,   {Umemura} M.,  2012, \mn@doi [\apj]
  {10.1088/0004-637X/757/2/136}, \href
  {https://ui.adsabs.harvard.edu/abs/2012ApJ...757..136W} {757, 136}

\bibitem[\protect\citeauthoryear{{Williams} et~al.,}{{Williams}
  et~al.}{2018}]{2018MNRAS.475.3429W}
{Williams} W.~L.,  et~al., 2018, \mn@doi [\mnras] {10.1093/mnras/sty026}, \href
  {https://ui.adsabs.harvard.edu/abs/2018MNRAS.475.3429W} {475, 3429}

\bibitem[\protect\citeauthoryear{{Wright}}{{Wright}}{2006}]{2006PASP..118.1711W}
{Wright} E.~L.,  2006, \mn@doi [\pasp] {10.1086/510102}, \href
  {https://ui.adsabs.harvard.edu/abs/2006PASP..118.1711W} {118, 1711}

\bibitem[\protect\citeauthoryear{{Yuan} \& {Wang}}{{Yuan} \&
  {Wang}}{2012}]{2012ApJ...744...84Y}
{Yuan} Z.,  {Wang} J.,  2012, \mn@doi [\apj] {10.1088/0004-637X/744/2/84},
  \href {https://ui.adsabs.harvard.edu/abs/2012ApJ...744...84Y} {744, 84}

\bibitem[\protect\citeauthoryear{{Zaja{\v{c}}ek} et~al.,}{{Zaja{\v{c}}ek}
  et~al.}{2019}]{2019A&A...630A..83Z}
{Zaja{\v{c}}ek} M.,  et~al., 2019, \mn@doi [\aap]
  {10.1051/0004-6361/201833388}, \href
  {https://ui.adsabs.harvard.edu/abs/2019A&A...630A..83Z} {630, A83}

\bibitem[\protect\citeauthoryear{{de Gasperin}, {Intema}  \& {Frail}}{{de
  Gasperin} et~al.}{2018}]{2018MNRAS.474.5008D}
{de Gasperin} F.,  {Intema} H.~T.,   {Frail} D.~A.,  2018, \mn@doi [\mnras]
  {10.1093/mnras/stx3125}, \href
  {https://ui.adsabs.harvard.edu/abs/2018MNRAS.474.5008D} {474, 5008}

\bibitem[\protect\citeauthoryear{{van Breugel}, {De Breuck}, {Stanford},
  {Stern}, {R{\"o}ttgering}  \& {Miley}}{{van Breugel}
  et~al.}{1999}]{1999ApJ...518L..61V}
{van Breugel} W.,  {De Breuck} C.,  {Stanford} S.~A.,  {Stern} D.,
  {R{\"o}ttgering} H.,   {Miley} G.,  1999, \mn@doi [\apjl] {10.1086/312080},
  \href {https://ui.adsabs.harvard.edu/abs/1999ApJ...518L..61V} {518, L61}

\bibitem[\protect\citeauthoryear{{van Weeren}, {de Gasperin}, {Akamatsu},
  {Br{\"u}ggen}, {Feretti}, {Kang}, {Stroe}  \& {Zandanel}}{{van Weeren}
  et~al.}{2019}]{2019SSRv..215...16V}
{van Weeren} R.~J.,  {de Gasperin} F.,  {Akamatsu} H.,  {Br{\"u}ggen} M.,
  {Feretti} L.,  {Kang} H.,  {Stroe} A.,   {Zandanel} F.,  2019, \mn@doi [\ssr]
  {10.1007/s11214-019-0584-z}, \href
  {https://ui.adsabs.harvard.edu/abs/2019SSRv..215...16V} {215, 16}

\makeatother
\end{thebibliography}

\appendix
\section{Fitted Parameters of Synchrofit}\label{app:synchrofit}
\begin{figure*}
\centering
\subfigure[SE hotspot with ALMA data, starts from 74 MHz.]{
\includegraphics[width=0.48\textwidth]{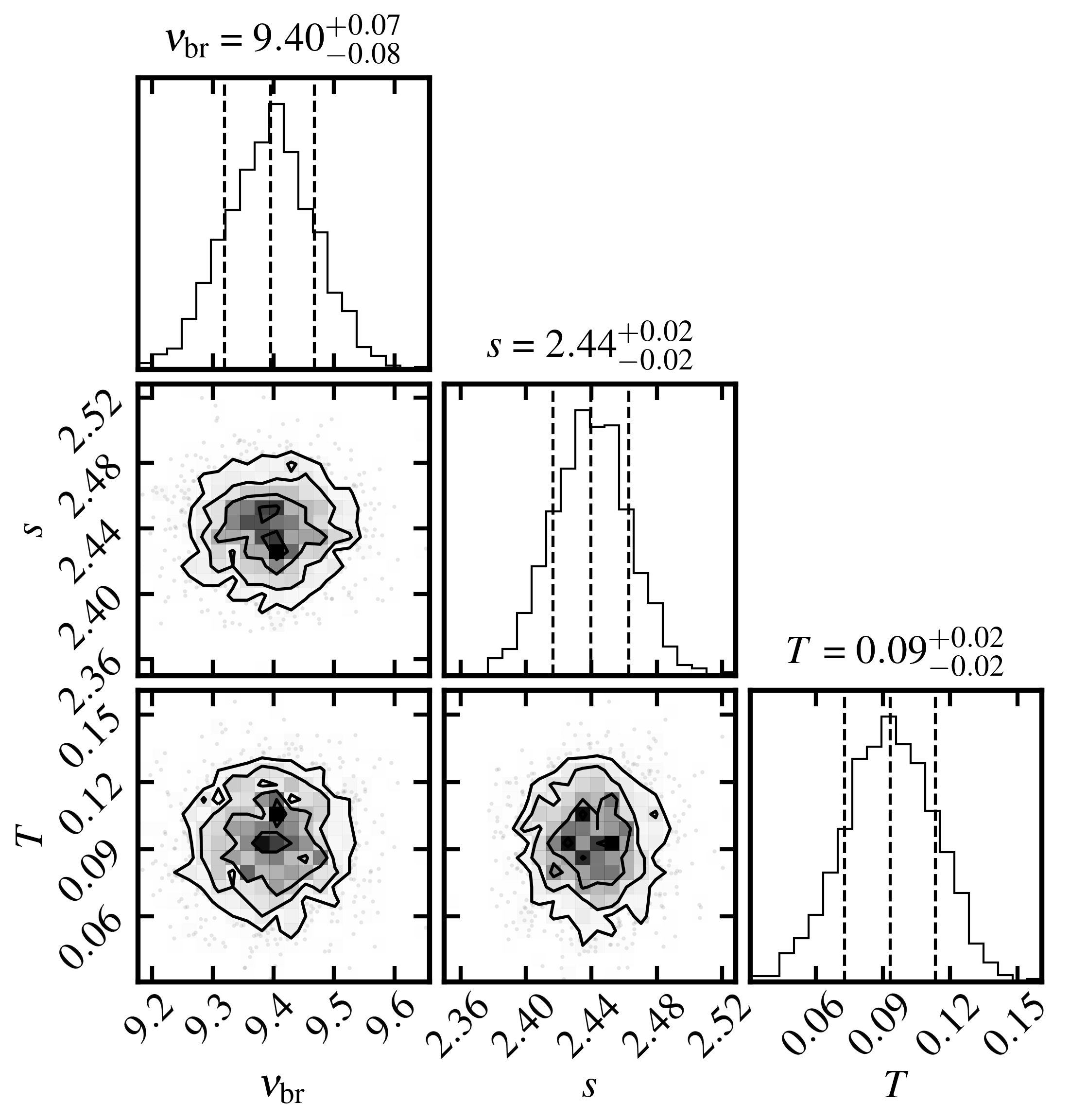}}
\hfill
\subfigure[SE hotspot with ALMA data, starts from 147 MHz.]{
\includegraphics[width=0.48\textwidth]{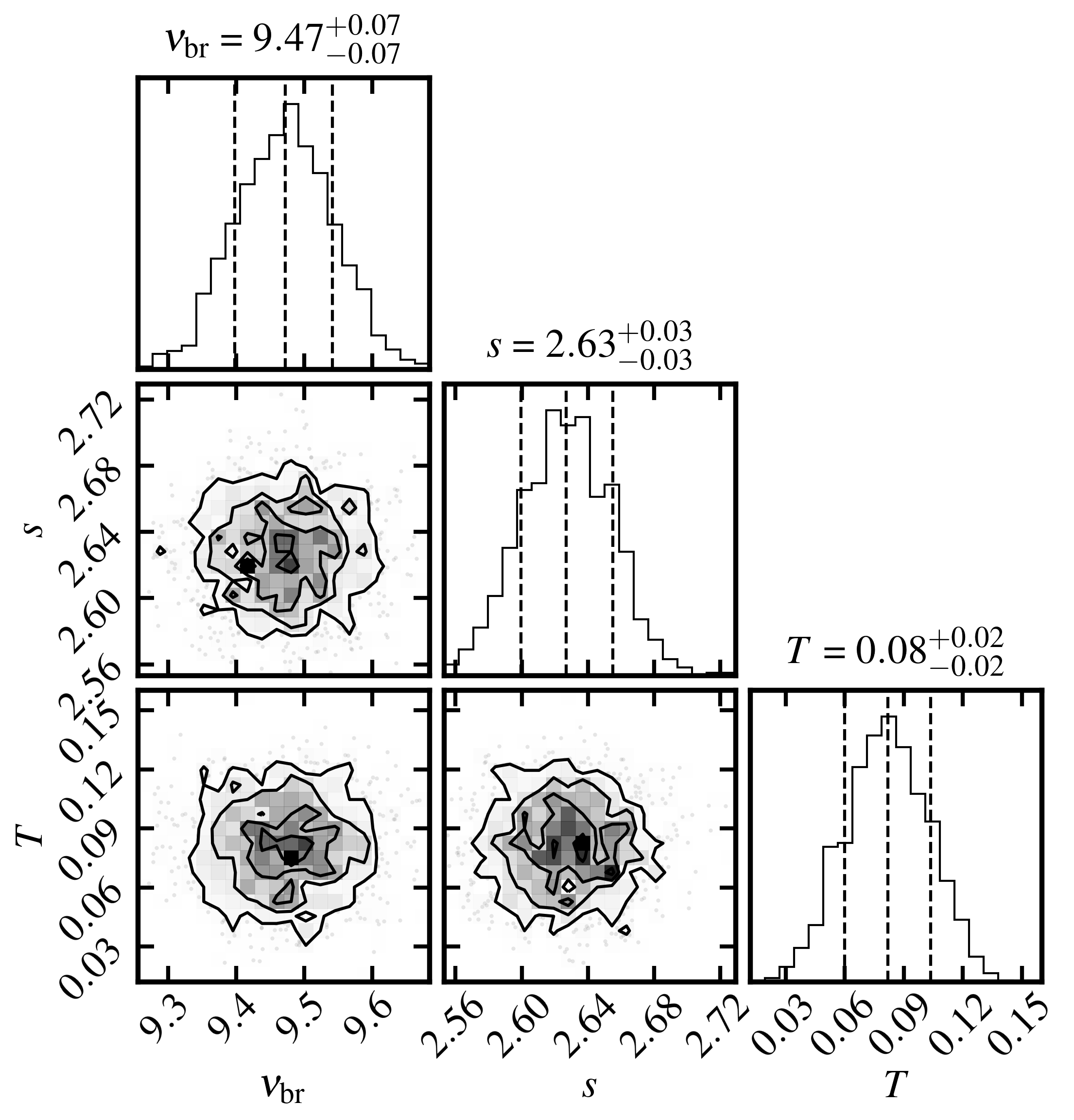}}
\caption{\label{fig:corner plots for synchrofit} 
Corner plots of the fitted parameters for comparisons between fittings with and without 74 MHz data.}
\end{figure*}

We show the corner plots of the break frequency $\nu_\mathrm{br}$, injection index $s$ (the injection spectral index is calculated by $\alpha_\text{inj}=(1-s)/2$), and remnant fraction $T$ fitted by \textsc{synchrofit} in Fig.~\ref{fig:corner plots for synchrofit}.
For comparison, we show the fittings with and without 74 MHz data in the left and right columns, respectively.
A shallower slope from 147 MHz towards 74 MHz has no significant impact on the fitted free parameters.

\begin{figure*}
\centering
\subfigure[Fitting of the NW hotspot assuming a flux ratio of 6:1]{
\includegraphics[width=0.48\textwidth]{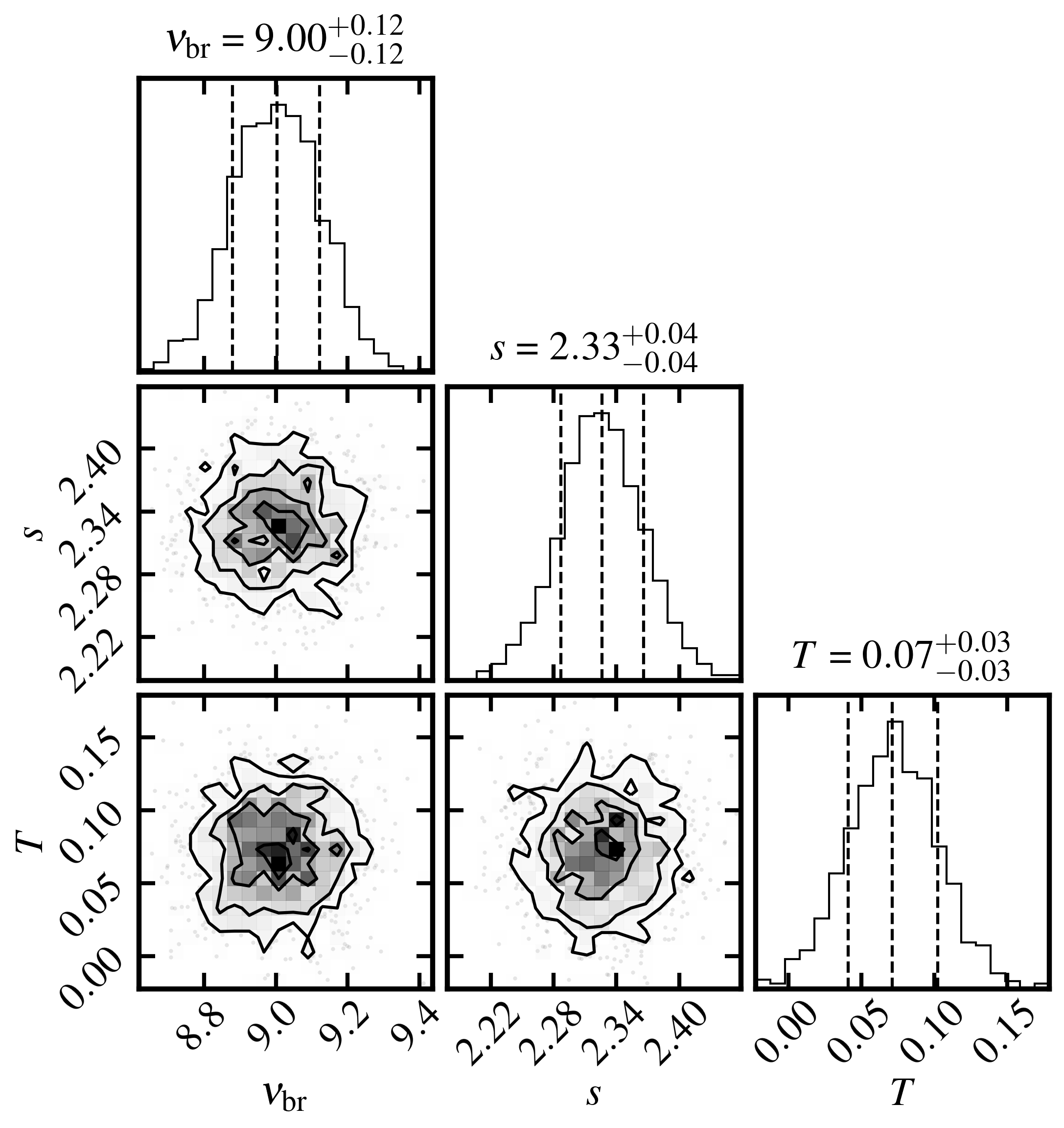}}
\hfill
\subfigure[Fitting of the NW hotspot assuming a flux ratio of 6:1. Only flux densities at 1.4, 4.7, 8.2, and 44 GHz are used.]{
\includegraphics[width=0.48\textwidth]{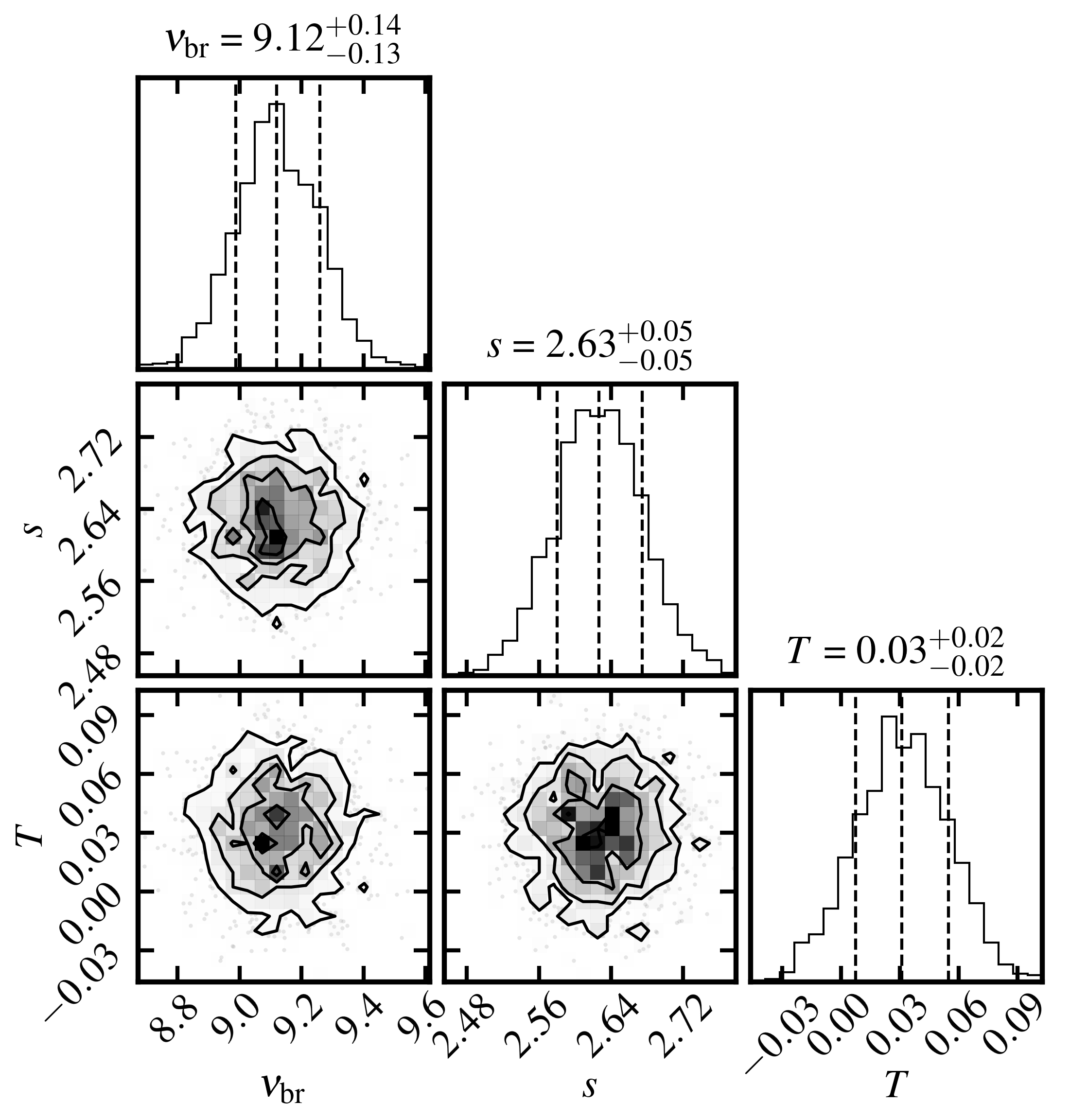}}
\caption{\label{fig:synchrofit for NW 6to1}
Corner plots of the fitted parameters of the NW hot spot for comparisons between fittings using 6 and 4 data points, respectively.
Although there is a relatively large difference between the injection indices, the break frequency and remnant fraction do not significantly differ.
These fitting results, compared with those presented in the main text, do not contradict the conclusions that the AGN is young and may have transient or intermittent activities.
We do note that because the limited number of data points used in the right one dilutes the curvature of the observed SED, a smaller remnant fraction is returned.}
\end{figure*}

\begin{figure*}
\centering
\subfigure[SE hotspot without ALMA data, assuming a flux ratio of 6:1.]{
\includegraphics[width=0.48\textwidth]{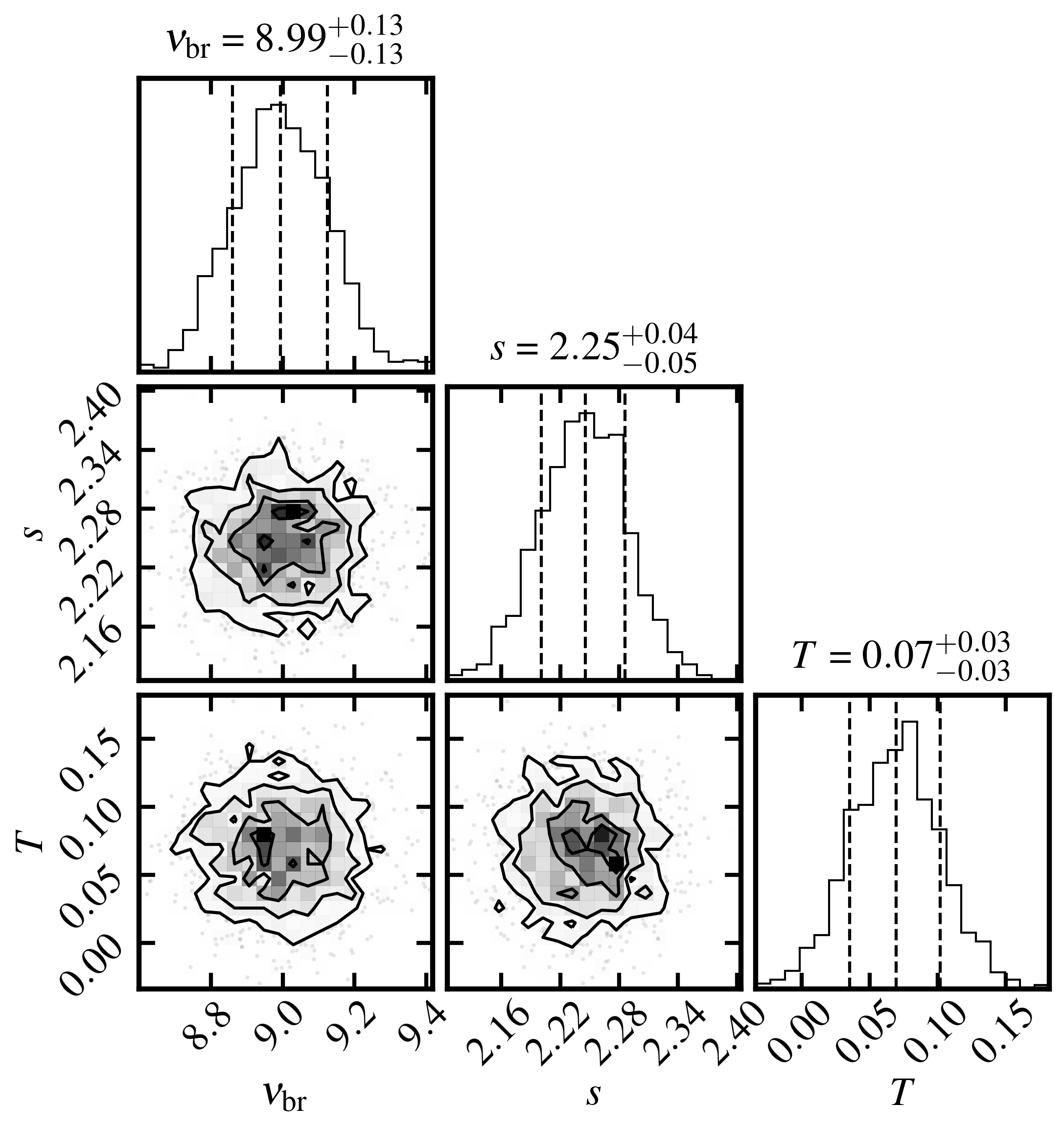}}
\hfill
\subfigure[SE hotspot with ALMA data, assuming a flux ratio of 6:1.]{
\includegraphics[width=0.48\textwidth]{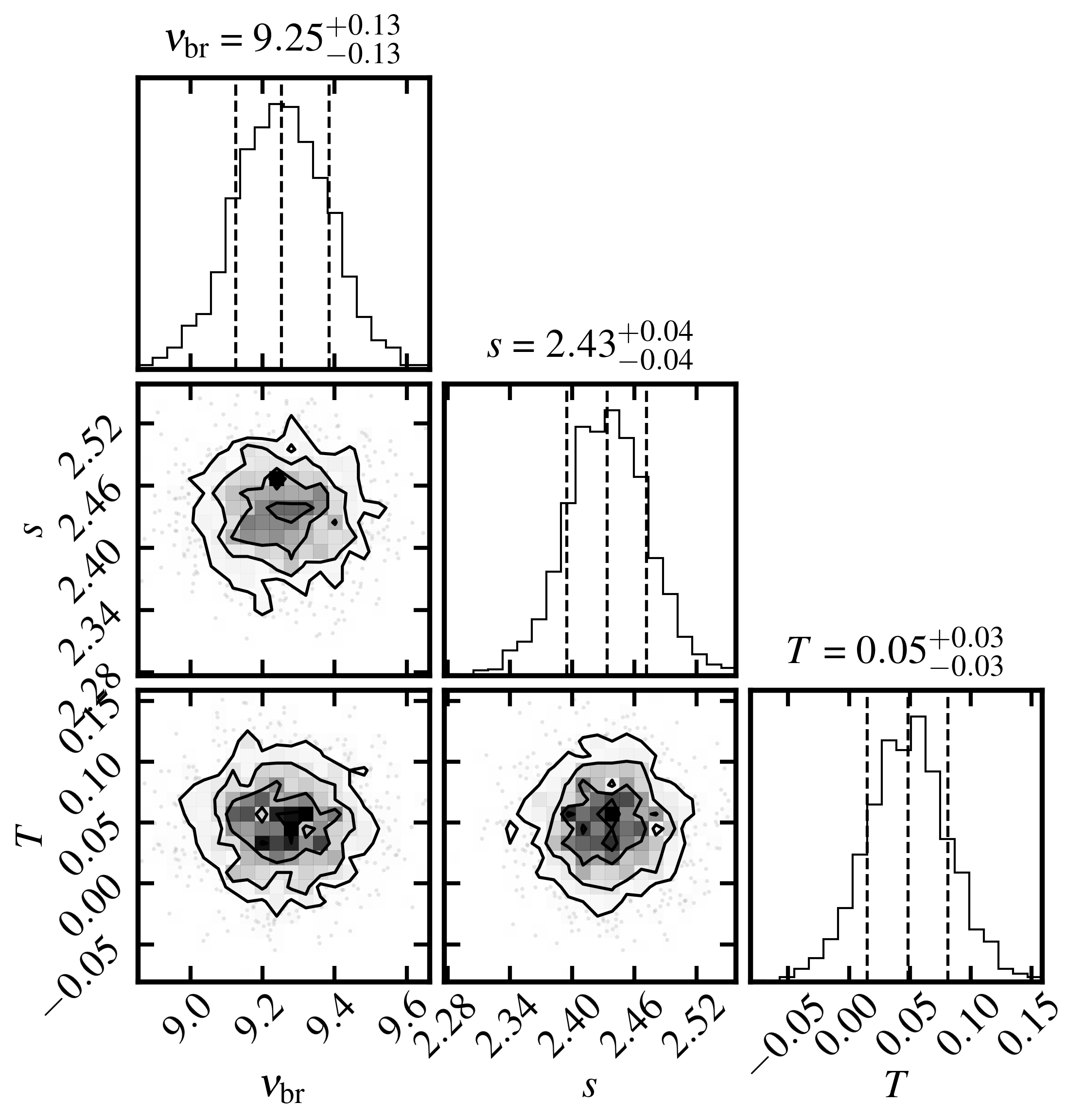}}
\caption{\label{fig:synchrofit for SE 6to1}
Corner plots of the fitted parameters of the SE hot spot for comparisons between fittings with and without ALMA 237 GHz data, assuming a flux ratio of 6:1 and subtracting possible contaminations from the low-resolution observations.
The injection index fitted without ALMA data, similar to the case of a flux ratio of 15:1 without considering contamination, is larger than that of the fitting with ALMA data.
Compared with the fittings presented in the main text, the break frequency can slightly migrate towards lower frequencies and there is a shorter time-scale of the quiescent phase.
However, these differences do not alter the conclusions we make in the main texts, that is, the AGN is young and may have transient or intermittent activities.}
\end{figure*}

\begin{figure*}
\begin{center}
\includegraphics[width=\textwidth]{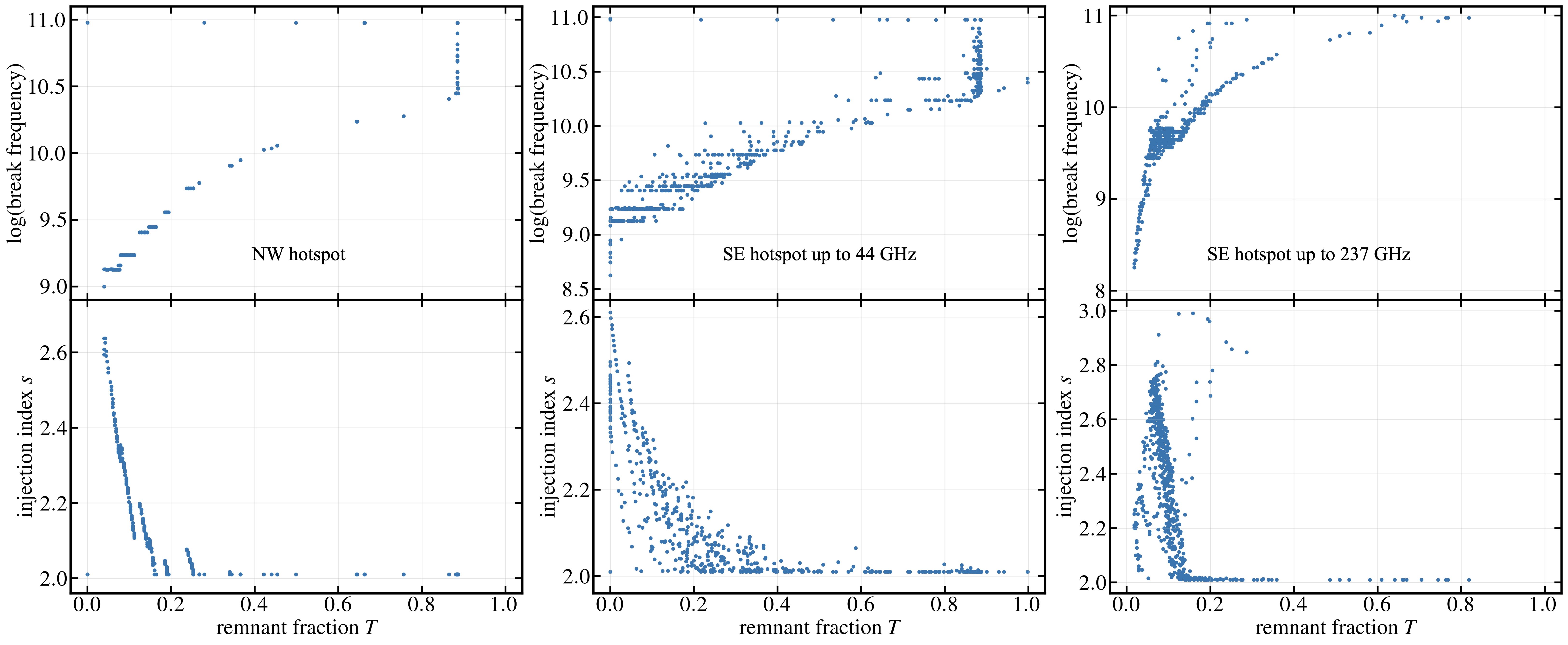}
\caption{Fitting results of NW and SE hotspot assuming the fractional contamination of diffuse radio emission ranging from 1 to 55 per cent by a step of 1 per cent and flux ratios ranging from 4 to 15 by a step of 1 to estimate the low-frequency flux densities.}
\label{fig:synchrofit statistics}
\end{center}
\end{figure*}

To investigate the impact of diffuse radio emissions on the fitting results and their implications for the young AGN and transient or intermittent radio activities, we performed more fittings assuming a flux ratio of $\tnr{S}{SE}:\tnr{S}{NW}=6:1$.
We further assume that the diffuse radio emission accounts for $\sim$50 per cent of the total observed flux densities at  GHz.
And at 4.7 and 8.2 GHz, the flux density of the SE hot spot is then six times the NW one.
The corresponding fitting results are shown in Fig.~\ref{fig:synchrofit for NW 6to1} and Fig.~\ref{fig:synchrofit for SE 6to1}.
We have also performed a fitting using only 4 data points following the same flux ratio as above to test whether the algorithm can still constrain the physical parameters.
The corresponding fitting result in shown in the right panel of Fig.~\ref{fig:synchrofit for NW 6to1}.
We further perform fittings assuming that the fraction of lobe contamination ranges from 1 to 55 per cent by a step of 1, and the flux density ratio varies from 4 to 15 by a step of 1 at low frequencies.
The statistics of the fitted parameters are shown in Fig.~\ref{fig:synchrofit statistics}.

\begin{figure*}
\centering
\subfigure[NW hotspot, starts from 74 MHz.]{
\includegraphics[width=0.48\textwidth]{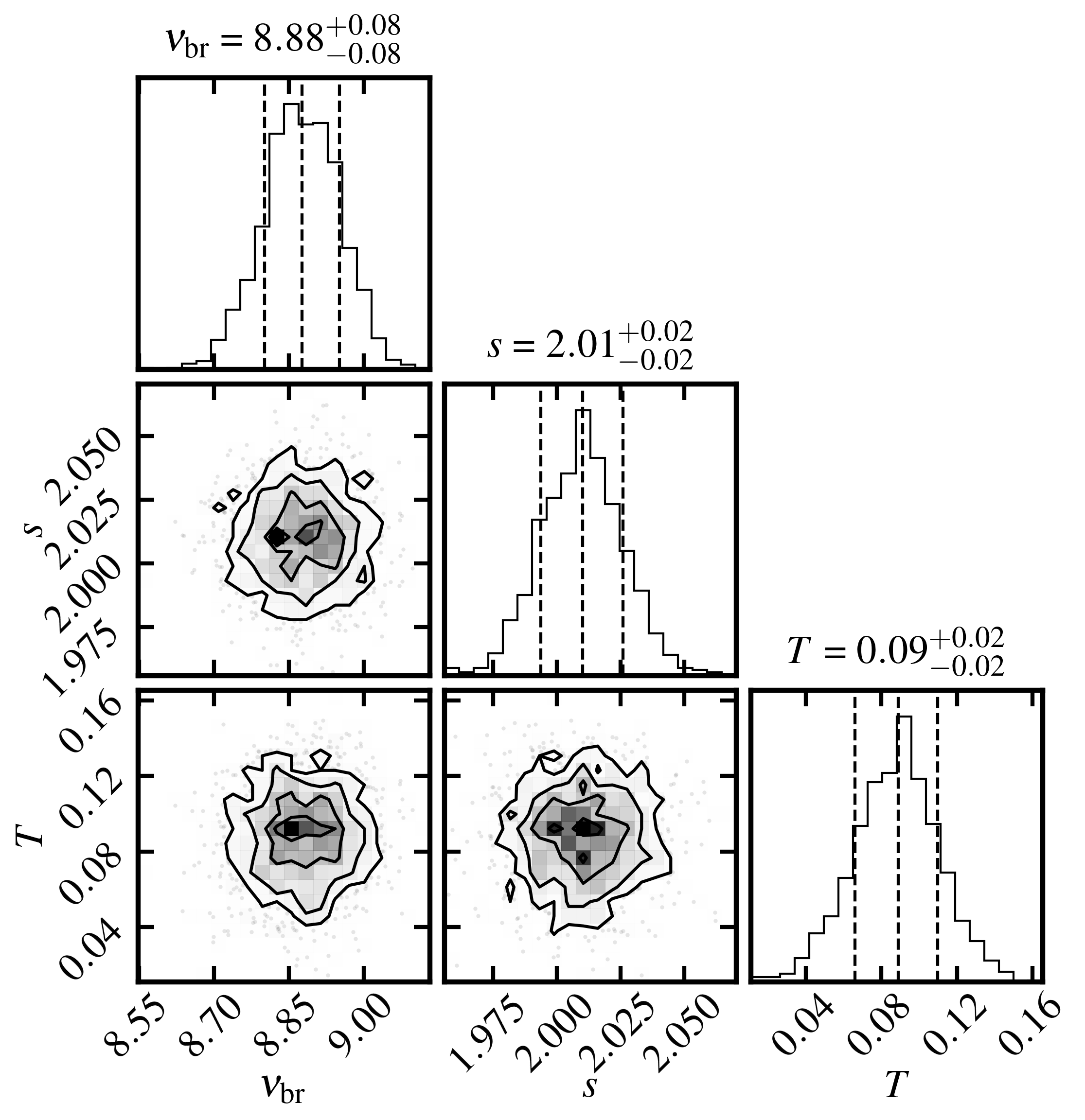}}
\hfill
\subfigure[NW hotspot, starts from 147 MHz.]{
\includegraphics[width=0.48\textwidth]{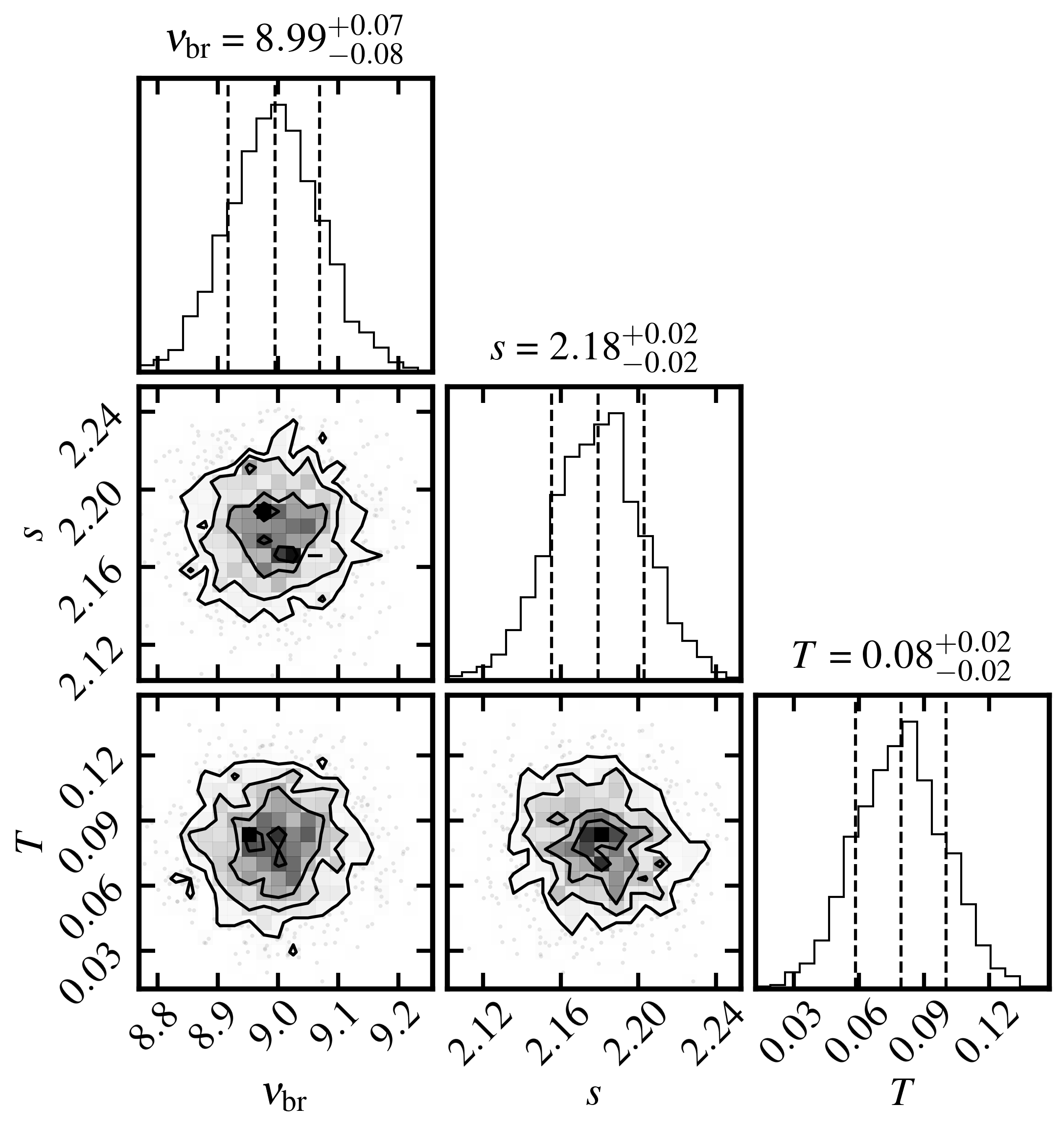}}
\qquad
\subfigure[SE hotspot without ALMA data, starts from 74 MHz.]{
\includegraphics[width=0.48\textwidth]{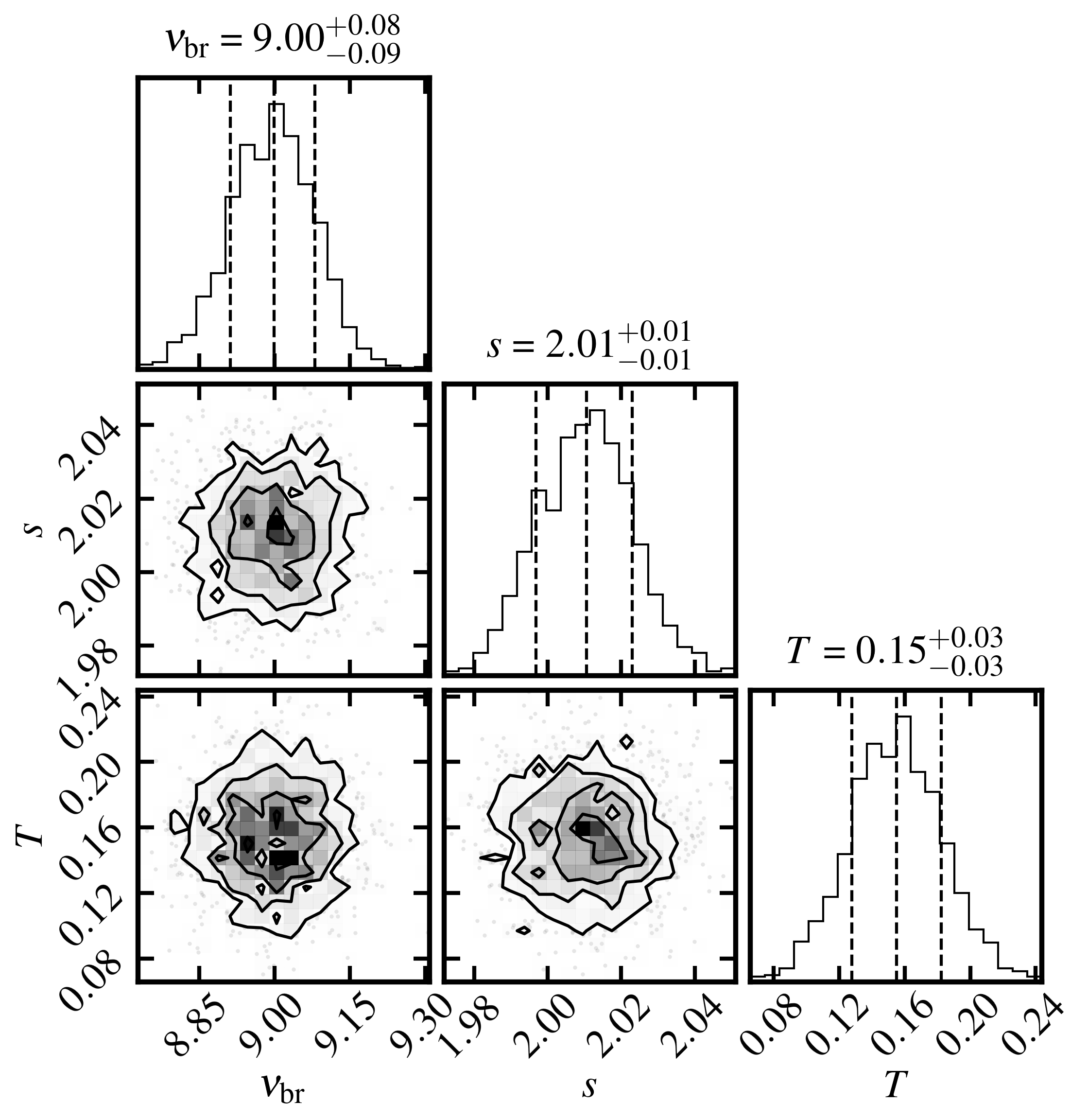}}
\hfill
\subfigure[SE hotspot without ALMA data, starts from 147 MHz.]{
\includegraphics[width=0.48\textwidth]{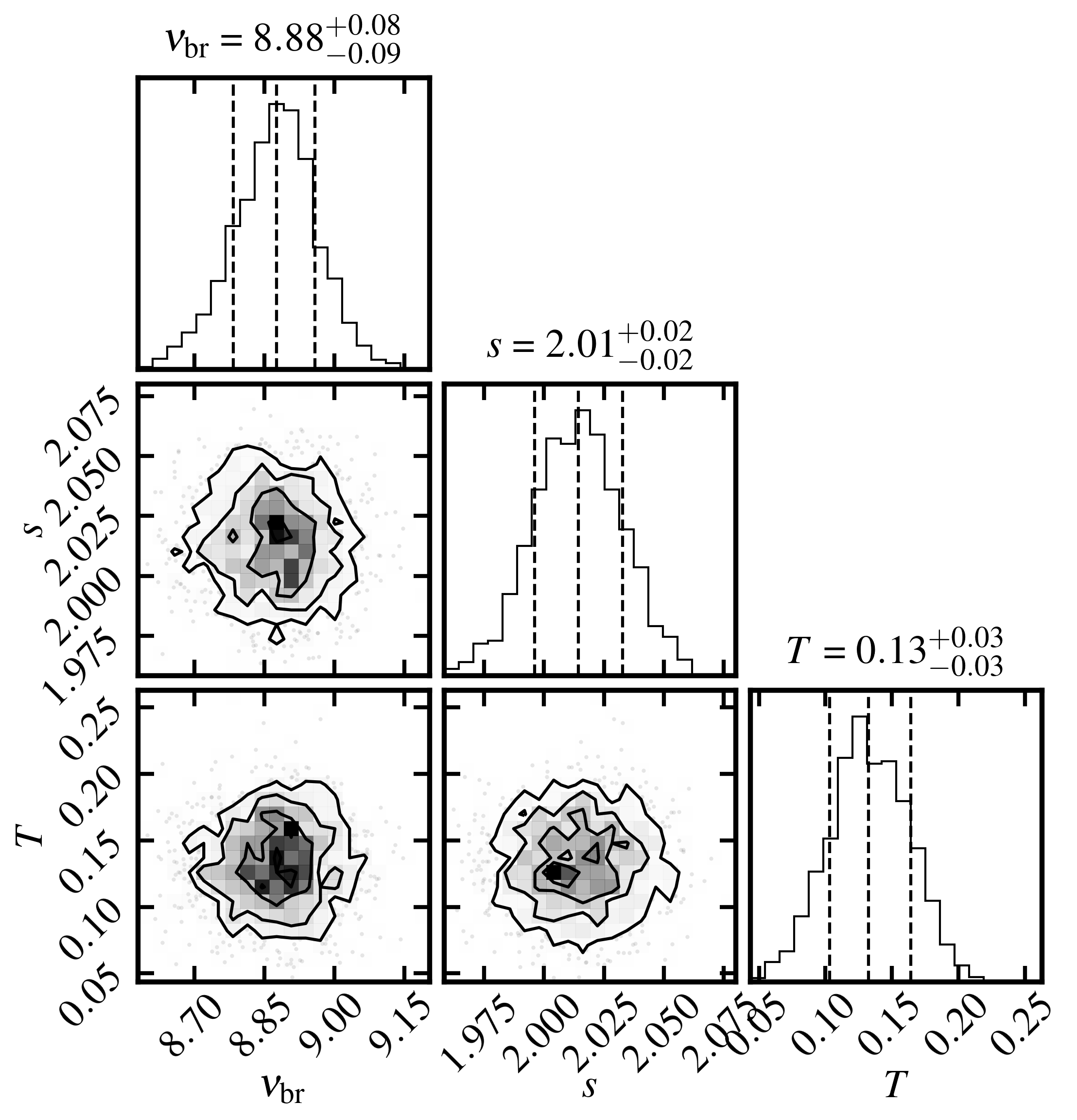}}
\end{figure*}

% Don't change these lines
\bsp	% typesetting comment
\label{lastpage}
\end{document}